\documentclass[reprint,twocolumn,secnumarabic,amssymb, aps, prl]{revtex4-1}

\usepackage{graphicx}
\usepackage{gensymb}
\usepackage{amsmath}
\usepackage[colorlinks,linkcolor=black,citecolor=black,urlcolor=black,filecolor=black]{hyperref}
\usepackage{wrapfig}
\begin{document}

\title{Extending the Quantum Coherence of a Near-Surface Qubit by Coherently Driving the Paramagnetic Surface Environment}

\author{Dolev Bluvstein, Zhiran Zhang, Claire A. McLellan, Nicolas R. Williams, and Ania C. Bleszynski Jayich}
\affiliation{Department of Physics, University of California, Santa Barbara, California 93106, USA}

\begin{abstract}
Surfaces enable useful functionalities for quantum systems, \textit{e.g.} as interfaces to sensing targets, but often result in surface-induced decoherence where unpaired electron spins are common culprits. Here we show that the coherence time of a near-surface qubit is increased by coherent radio-frequency driving of surface electron spins, where we use a diamond nitrogen-vacancy (NV) center as a model qubit. This technique is complementary to other methods of suppressing decoherence, and importantly, requires no additional materials processing or control of the qubit. Further, by combining driving with the increased magnetic susceptibility of the double-quantum basis we realize an overall fivefold sensitivity enhancement in NV magnetometry. Informed by our results, we discuss a path toward relaxation-limited coherence times for near-surface NV centers. The surface spin driving technique presented here is broadly applicable to a wide variety of qubit platforms afflicted by surface-induced decoherence.
\end{abstract}

\maketitle
Decoherence of quantum systems near surfaces is an outstanding challenge that has not been met. Whereas many quantum systems benefit from a high degree of environmental control and regularity, such as atoms trapped in vacuum far from surfaces and atomic-scale defects buried deep in a bulk crystal, interfaces can add important functionality towards scalability, transduction, and networking. However, interfaces also add an uncontrolled element to the qubit's environment. Surfaces in particular are inevitable for superconducting qubits, nanomechanical resonators, adatom qubits, atom and ion chip traps, and for high spatial resolution sensing, but frequently host sources of decoherence \cite{Wang2015,MacCabe2019,Baumann2015,Lin2004,Turchette2000,Myers2014}. In particular, fluctuating magnetic fields from surface electron spins are implicated as a major source of decoherence for qubits near surfaces \cite{Schenkel2006,Sendelbach2008,Bluhm2009,Ofori-Okai2012b,Rosskopf2014,Romach2015,DeGraaf2017}.

The negatively charged nitrogen-vacancy (NV) center in diamond is a renowned, model example of a solid-state qubit \cite{Doherty2013a}. Near-surface NV centers are used as versatile, high spatial resolution quantum sensors \cite{Staudacher2013a,Gross2017,Casola2018,Ariyaratne2018b} and are also useful for storing, processing, and transferring quantum information in hybrid quantum systems \cite{Faraon2011a,Cai2013a,Golter2016}. The NV coherence time is a key parameter in these quantum applications as it directly limits the possible storage and processing time as well as the achievable sensitivity in sensing \cite{Lovchinsky2016a,Degen2017}. For these applications it is vital that the NV center reside in close proximity to the diamond surface, as the NV depth determines the coupling strength to other quantum elements, as well as the signal strength and spatial resolution in imaging \cite{Taylor2008b,Pelliccione2016b}. However, as for a wide variety of other quantum systems, decoherence associated with the surface is a key obstacle in NV-based technologies \cite{Romach2015,Kim2015,Myers2017b}.

Several approaches are commonly taken to mitigate decoherence of near-surface qubits. Surface engineering is a direct, materials-based approach, which often involves preparing the surface with a host of material-specific protocols \cite{Tisler2009,Hite2013,Quintana2014,Kumar2016a}. Although surface engineering is a promising approach that directly targets the source of the problem, discovering the correct protocols can require painstaking characterization and trial-and-error; to date, state-of-the-art techniques have not succeeded in completely eliminating the decoherence sources.
Further, the surface can degrade in time after the initial preparation, due to spontaneous chemical or structural changes as well as accumulated surface adsorbates \cite{McGuirk2004,Daniilidis2011,Hite2012,Bluvstein2019}. Quantum control of the near-surface qubit is another approach to mitigating decoherence. Dynamical decoupling is a primary example, where by fast rotation the qubit is made insensitive to slower frequencies of fluctuations \cite{Du2009,DeLange2010,Bylander2011}. One can also use qubit states with ``clock transitions'' that are insensitive to specific perturbations such as magnetic fields \cite{Koch2007,Wolfowicz2013,Zadrozny2017,Barfuss2018}. Although they can increase coherence time, these techniques significantly constrain the qubit and its applications, rendering it insensitive to signals that are similar in nature to the noise being decoupled. An active approach that directly addresses only the offending source and leaves the qubit unconstrained, without painstaking materials processing or susceptibility to surface degradation, is a promising path forward.

\begin{figure}
\includegraphics[width=86mm]{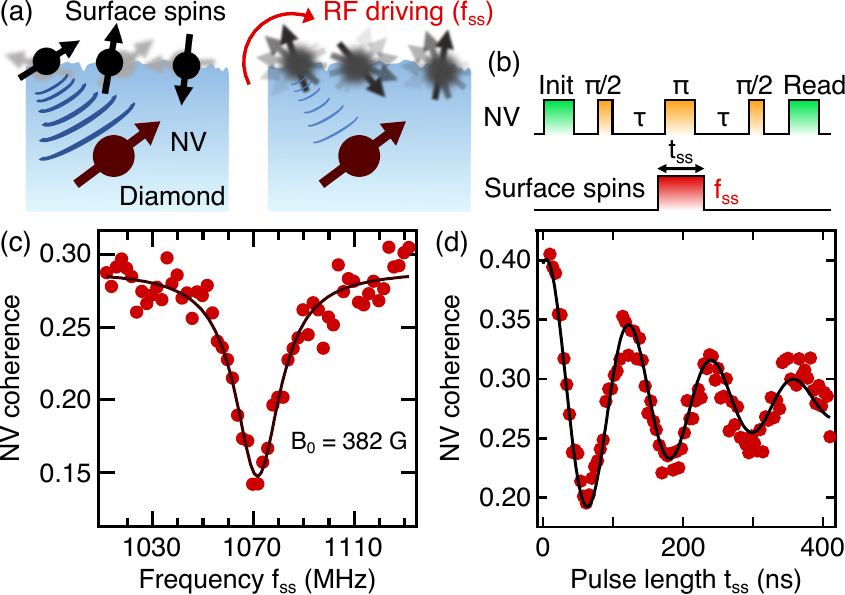}
\caption{(a) Schematic of experiment. Near-surface NV centers are dephased by fluctuating magnetic fields (blue contours) from surface electronic spins. Driving the surface spins can suppress NV dephasing. (b) Pulse sequence used in (c) and (d) for probing surface spins. Microwave pulses (yellow and red) are used for spin control and green illumination pulses are used for initializing and reading the NV spin. When an on-resonance $\pi$ pulse inverts the surface spins, their quasistatic magnetic field is recoupled and induces NV dephasing. (c) NV coherence as a function of pulse frequency $f_\text{SS}$, showing a resonance corresponding to $g=2$ electronic spins. A $\pi$ pulse of duration $t_\text{SS}$ = 108 ns is used. Black curve is a Lorentzian fit. (d) NV coherence as a function of pulse length $t_{\text{SS}}$ with $f_\text{SS} = 1071$ MHz. Black curve is a fit to exponentially damped Rabi oscillations with $T_{2,\text{Rabi}} = 200(10)$ ns. 2$\tau$ is 20 $\mu$s in (c) and (d) and the NV depth is 7.5(3) nm.
}
\label{fig1}
\end{figure}

In this work we demonstrate that coherent radio-frequency driving of surface electron spins removes their decohering effect on nearby qubits. Here we use a diamond NV center as a model near-surface qubit. Importantly, the qubit coherence time increases without any additional materials processing or manipulation of the qubit. The physical principle of this approach is that by driving the surface spins sufficiently fast, their interaction with the qubit averages away [Fig.~\ref{fig1}(a)], analogous to the phenomenon of motional narrowing \cite{Slichter1990}. With NV centers specifically, we realize a fivefold enhancement in measurement sensitivity by combining surface spin driving with the increased magnetic susceptibility of the double-quantum basis. We find the coherence extension from driving is robust among individual NVs. The techniques here are broadly applicable to qubits affected by paramagnetic environments and are directly complementary to other methods of suppressing decoherence. Lastly we discuss a path toward realizing relaxation-limited coherence times for near-surface NV centers by combining surface-spin driving with existing materials processing techniques.

The experimental setup consists of a home-built, room-temperature confocal microscope for optically addressing individual near-surface NV centers in a single-crystal diamond plate. Three separate radio-frequency (RF) signal generators are used for controlling the electronic spin state of the NV spin qutrit, formed by $|m_s=0\rangle_{\text{NV}}$ and $|m_s=\pm1\rangle_{\text{NV}}$, as well as the electronic spin state of surface spin qubits, formed by $|\uparrow \rangle_{\text{SS}}$ and $|\downarrow\rangle_{\text{SS}}$. A single microwave waveguide patterned on the diamond is used to deliver all RF signals.

The diamond substrate used here is prepared by chemical vapor deposition growth of a 50-nm-thick 99.99\% $^{12}$C layer onto an Element Six electronic grade (100) diamond. NV centers are then formed by 4 keV $^{14}$N ion implantation with a dosage of 5.2$\times 10^{10}$ ions/cm$^2$ into the diamond plate, followed by annealing in vacuum at 850 $\degree$C for 2.5 h. The surface is then tri-acid cleaned and annealed in an oxygen atmosphere \cite{Sangtawesin2018} (see Supplemental Material Note 1 for further details \footnote{See the Supplemental Material for supporting details on sample information, surface spin density, RF control, the AC Stark effect, and DQ and SQ susceptibilities, which includes Refs.~\cite{Toyli2010,DeWit2018,Friedrich2017}}). The NV centers' depths are experimentally measured via surface proton NMR and range between $\sim$4 and 17 nm \cite{Pham2016}. 

The electron spins at the diamond surface and their interactions with single, shallow NV centers can be probed by a double electron electron resonance (DEER) measurement sequence [Fig.~\ref{fig1}(b)], in which the NV center serves as a local, optically addressable readout for a small number of proximal dark spins \cite{Mamin2012,Grinolds2014,Sushkov2014}. Specifically, a Hahn echo sequence is performed on the NV, in which a resonant $\pi$ pulse on the $|m_s = 0 \rightarrow -1 \rangle_{\text{NV}}$ transition decouples the NV center from slowly varying environmental fluctuations that lead to decoherence; simultaneously, a coherent microwave pulse tuned to the surface spins (frequency $f_\text{SS}$) selectively recouples their quasistatic contribution and leads to NV decoherence. Therefore, in this DEER measurement, the NV coherence serves as a readout for the surface spins.

In Figs.~\ref{fig1}(c) and \ref{fig1}(d) we probe the surface spin frequency- and time-domain response to RF fields using the DEER measurement sequence in Fig.~\ref{fig1}(b). Here we use spin-dependent photoluminescence to measure NV coherence in the single-quantum basis $\{|m_s=0\rangle,|m_s=-1\rangle\}_{\text{NV}}$ at a fixed echo time $2\tau$. The plotted NV coherence is defined as the difference between the populations of the $|m_s=0\rangle_{\text{NV}}$ and $|m_s=-1\rangle_{\text{NV}}$ states after the final $\pi/2$ pulse in Fig.~\ref{fig1}(b) (neglecting the non-unity spin polarization \cite{Note1}). Figure~\ref{fig1}(c) shows the surface spin electron spin resonance spectrum, obtained by measuring NV coherence as the frequency $f_{\text{SS}}$ of a fixed-duration surface spin pulse is varied. A clear spin resonance is seen at 1071 MHz, which is the expected resonance for $g=2$ spins at the applied magnetic field of $B_0$ = 382 G. Figure~\ref{fig1}(d) shows time-domain Rabi oscillations of the surface spins, obtained by fixing $f_\text{SS}=1071$ MHz and varying the duration $t_\text{SS}$ of the surface spin pulse, demonstrating our ability to address and coherently control these $g=2$ surface spins.

\begin{figure}
\includegraphics[width=85.7mm]{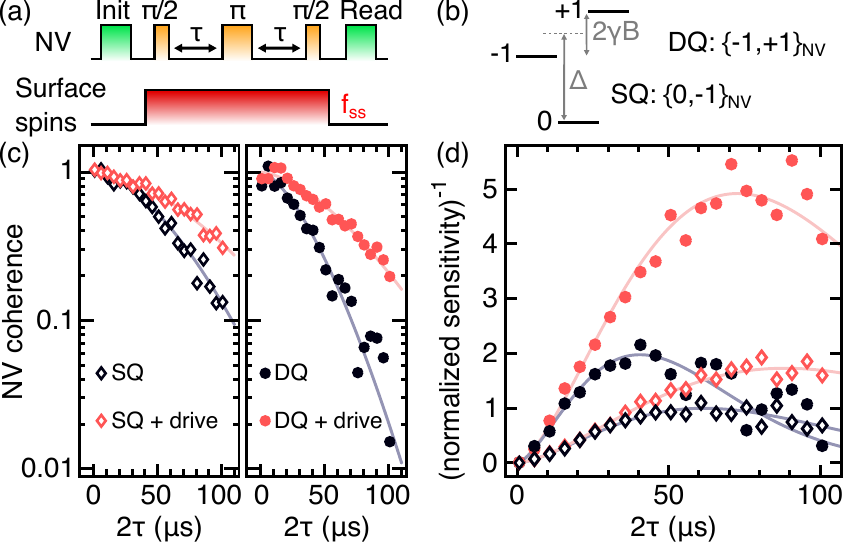}
\caption{(a) NV Hahn echo sequence with continuous surface spin drive, used in (c). (b) NV ground state spin-1 energy level diagram. The double-quantum (DQ) basis $\{-1,+1\}_\text{NV}$ is insensitive to common-mode $\Delta$ fluctuations and has greater magnetic field $B$ susceptibility than the single-quantum (SQ) basis $\{0,-1\}_\text{NV}$. 
(c) NV Hahn echo $T_2$ coherence decay measured in the SQ (left) and DQ (right) bases, with and without surface spin driving, showing that driving extends coherence. Data are fit to $\exp(-(2\tau/T_2)^n)$ with fitted $n \approx 1.6$. $T_{2,\text{SQ}} = 65(2) ~\mu$s (dark diamonds), $T_{2,\text{SQ+drive}} = 94(2) ~\mu$s (red diamonds), $T_{2,\text{DQ}} = 41(3) ~\mu$s (dark circles), $T_{2,\text{DQ+drive}} = 75(3) ~\mu$s (red circles). Surface spin $\Omega_\text{Rabi}/2\pi = 10$ MHz and the NV depth is 12.8(3) nm. (d) NV center's inverse sensitivity to magnetic field variance, calculated for the data in (c) and normalized to the peak sensitivity for SQ without drive. A $5\times$ sensitivity enhancement is observed for the DQ + drive measurement.
}
\label{fig2}
\end{figure}

Having identified surface spins and established their coherent control, we now show that by coherently driving them we remove their decohering effect on a near-surface qubit. In Fig.~\ref{fig2}(c) we observe an increase in the NV Hahn echo $T_2$ in both the single-quantum (SQ) and double-quantum (DQ) bases by continuously driving the surface spins on-resonance throughout the echo sequence [sequence in Fig.~\ref{fig2}(a)]. Single-tone pulses from two separate RF generators are used to control the NV ground state spin triplet in either the SQ basis $\{0,-1\}_\text{NV}$ or DQ basis $\{-1,+1\}_\text{NV}$ (see Supplemental Material Note 3 for further details \cite{Note1}). As diagrammed in Fig.~\ref{fig1}(a), by coherently driving these spins sufficiently fast, their magnetic interaction with the NV averages away and surface-spin-induced dephasing is suppressed. For the NV in Fig.~\ref{fig2}, the SQ $T_2$ increases from 65(2) to 94(2) $\mu$s by driving and the DQ $T_2$ increases from 41(3) to 75(3) $\mu$s by driving.
Here we drive the surface spins at $\Omega_\text{Rabi}/2\pi = 10$ MHz, which is roughly 2$\times$ the half width at half maximum of the surface spin linewidth ($1/T_{2,\text{Rabi}} \approx 5$ MHz) and is much greater than the NV coupling strength to an individual surface spin ($\lesssim 10$ kHz). The coherence extension we observe is robust: by driving we observe a SQ $T_2$ increase of $>15\%$ for 11 out of 13 measured centers, and we observe up to a 140\% increase.

\begin{figure}
\includegraphics[width=86mm]{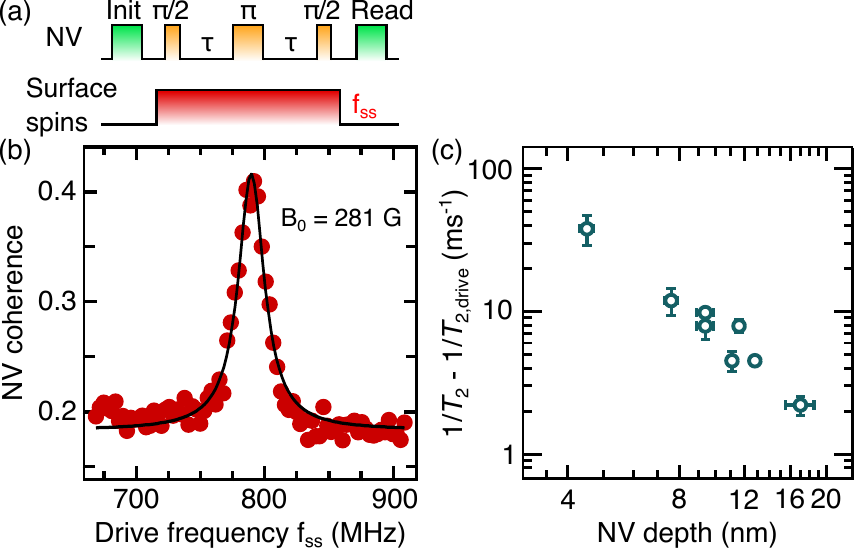}
\caption{NV coherence extension is due to resonant driving of $g=2$ electron spins at the diamond surface. (a) NV Hahn echo sequence with continuous surface spin drive, used in (b). (b) NV Hahn echo coherence in the SQ basis at fixed $\tau$ and varying $f_\text{SS}$, showing a resonance corresponding to $g=2$ spins. Black curve is a Lorentzian fit. Surface spin $\Omega_\text{Rabi}/2\pi = 7$ MHz, 2$\tau$ = 34 $\mu$s and the NV depth is 11.6(3) nm. (c) Decoupled SQ decoherence rate by driving at $g=2$ resonance, plotted as a function of NV depth. The decoupled rate is strongly anti-correlated with NV depth.
}
\label{fig3}
\end{figure}

Before further discussing the results from Fig.~\ref{fig2}, we briefly show that the observed $T_2$ increase is due to resonant driving of $g=2$ electron spins and that these spins reside at the diamond surface. Figure~\ref{fig3}(b) plots measurements of NV coherence at fixed echo time $2\tau$ while continuously driving at variable frequency $f_\text{SS}$ [Fig.~\ref{fig3}(a)]. Sweeping $f_\text{SS}$ across the expected $g=2$ resonance (788 MHz at the applied $B_0$ = 281 G), we observe a clear increase in the NV coherence, indicating that the coherence extension from continuous drive is due to resonant driving of $g=2$ spins. To confirm these spins are at the surface, in Fig.~\ref{fig3}(c) we plot the decoupled SQ decoherence rate $\Gamma_\text{decoupled} = 1/T_{2} - 1/T_{2,\text{drive}}$ as a function of NV depth, measured separately on eight individual centers with depths ranging from 4 to 17 nm. We observe a strong anticorrelation between $\Gamma_\text{decoupled}$ and the NV depth, with over an order of magnitude change in $\Gamma_\text{decoupled}$ between the shallowest and deepest centers, demonstrating that the decoupled noise originates at the surface. We further note that on day-to-month timescales we observe order-unity variations in $\Gamma_\text{decoupled}$ measured on the same NV, further indicating that the decoupled decoherence originates from the surface.

We now turn to a discussion of the coherence extension in the single- and double-quantum bases shown in Fig.~\ref{fig2}. In the SQ and DQ bases, the NV has different susceptibilities to the various dephasing channels \cite{Mamin2014,Bauch2018}, which can be quantitatively understood by the NV's ground state spin Hamiltonian

\begin{equation}
    H_\text{NV} = \left(hD + d_{\|} \Pi_{\|}\right) S_z^2 + h \frac{\gamma}{2\pi}\boldsymbol{S} \cdot \boldsymbol{B} - \frac{d_{\perp} \Pi_{\perp}}{2} \left(S_+^2 + S_-^2\right),
    \label{Hamiltonian}
\end{equation}

where $h$ is Planck's constant, $D = 2.87$ GHz is the crystal-field splitting, $\boldsymbol{B}$ is the magnetic field, $\boldsymbol{S}$ is the spin-1 operator, $S_{\pm}$ are the spin raising and lowering operators, the $z$-axis points along the NV axis, $\gamma/2\pi = 2.8$ MHz/G is the NV gyromagnetic ratio, $d_{\|}/h = 0.35$ Hz$\cdot$cm/V and $d_{\perp}/h = 17$ Hz$\cdot$cm/V are the components of the NV's electric dipole parallel and perpendicular to the $z$-axis, and $\Pi_{\|}$ and $\Pi_{\perp}$ are the parallel and perpendicular components of the effective electric field, where $\boldsymbol{\Pi} = (\boldsymbol{E} + \boldsymbol{\sigma})$ has both electric field $\boldsymbol{E}$ and appropriately scaled strain $\boldsymbol{\sigma}$ terms. With an applied $\boldsymbol B = B_z \hat{z}$, where $(\gamma/2\pi) B_z \gg d_{\perp} \Pi_{\perp}/h$, the Hamiltonian yields SQ transition frequencies \cite{Kim2015,Bauch2018} given by

\begin{equation}
    f_{0\rightarrow\pm1} \approx D + d_{\|} \Pi_{\|}/h \pm \left(\frac{\gamma}{2\pi} B_z + \frac{1}{2} \frac{\left(d_{\perp} \Pi_{\perp}/h\right)^2}{(\gamma/2\pi) B_z}\right),
    \label{SQFrequencies}
\end{equation}

and a DQ transition frequency given by

\begin{equation}
    f_{-1\rightarrow+1} \approx 2 \left(\frac{\gamma}{2\pi} B_z + \frac{1}{2} \frac{\left(d_{\perp} \Pi_{\perp}/h\right)^2}{(\gamma/2\pi) B_z}\right).
    \label{DQFrequencies}
\end{equation}

The DQ basis has an effectively doubled gyromagnetic ratio, which has the positive effect of being more sensitive to magnetic signals but is traded off with an increased sensitivity to magnetic noise. A clear advantage of the DQ basis, however, is the elimination of noise from the common-mode $D + d_{\|} \Pi_{\|}/h$ terms. This advantage is borne out in Fig.~\ref{fig2}(c) where we observe $(T_{2,\text{DQ}} / T_{2,\text{SQ}})^n = 0.48(4)$ (where $n = 1.6$ is the exponential stretch factor), which is greater than 0.25, the expected value for purely magnetic noise (see Supplemental Material Note 5.1 \cite{Note1}), indicating the elimination of substantial common-mode noise \cite{Kim2015}.

Importantly, the advantage of operation in the DQ basis can be amplified by driving the surface spins to decouple a large portion of the magnetic environment, and in doing so, we achieve $T_{2,\text{DQ+drive}} > T_{2,\text{SQ}}$ [Fig.~\ref{fig2}(c)]. In result, we achieve greatly amplified sensitivity gains in magnetometry: the longer coherence time allows for a longer phase accumulation time from a magnetic signal, and the doubled gyromagnetic ratio results in a doubled phase accumulation rate \cite{Bauch2018}. To quantify these magnetometry improvements, we consider the sensitivity enhancements to incoherent AC signals, which is relevant in, \textit{e.g.}, noise detection of magnetic phases \cite{Hall2009,Schafer-Nolte2014a,Casola2018} or detection of the statistical polarization of 
precessing nuclear spins in nanoscale NMR \cite{Meriles2010,Staudacher2013a,Lovchinsky2016a}. In this case, the signal strength for the optically detected signal goes as $C(T) \langle \left(\delta \phi\right)^2\rangle$, where $\langle \left(\delta \phi\right)^2\rangle \sim B^2_{\text{rms}} \gamma_{\text{eff}}^2 T^2$ is the variance in accumulated phase, $T$ is the total phase accumulation time, $C(T)$ is the NV coherence, $B_{\text{rms}}$ is the root mean square of the magnetic field, and $\gamma_{\text{eff}}$ is the effective gyromagnetic ratio: $\gamma_{\text{eff}} = \gamma$ in the SQ basis and $\gamma_{\text{eff}} = 2 \gamma$ in the DQ basis. The inverse sensitivity (high inverse sensitivity corresponds to faster sensing) thus goes as (signal strength)$/\sqrt{T} \sim C(T)\gamma_{\text{eff}}^2 T^{3/2}$ \cite{Lovchinsky2016a}. 

Figure~\ref{fig2}(d) plots inverse sensitivity for the data measured in Fig.~\ref{fig2}(c), normalized to the peak sensitivity for the no-drive, SQ measurement; accordingly, the y-axis can be understood as a sensitivity enhancement. We observe a $2\times$ sensitivity enhancement in the no-drive DQ measurement, a $1.75\times$ enhancement in the SQ + drive measurement, and a $5\times$ enhancement in the DQ + drive measurement. This $5\times$ sensitivity enhancement corresponds to a $25\times$ measurement speed-up. We note that the SQ, no-drive $T_2$ measured here is similar to that reported for NVs of this depth (12.8(3) nm) produced by state-of-the-art materials techniques \cite{Myers2014,FavarodeOliveira2017a,Sangtawesin2018}; and with the enhanced collection from our diamond nanopillars, we estimate an a.c. magnetic field sensitivity of $\eta_\text{a.c.} = 10 ~\text{nT Hz}^{-1/2}$ for the DQ + drive Hahn echo \cite{Taylor2008b}.

We now highlight a path pushing toward $T_1$-limited coherence times for near-surface NVs. With the advances in this work, the suspected remaining limits to shallow NV coherence in our diamond sample are $\Pi_{\perp}$ fluctuations and bulk electron spins which produce $B$ fluctuations. These dephasing sources can be mitigated by existing techniques that are directly complementary to the methods used in this work. First, the $\frac{1}{2} \frac{\left(d_{\perp} \Pi_{\perp}/h\right)^2}{(\gamma/2\pi) B_z}$ term is suppressed with large magnetic fields. By comparing DQ and SQ coherence times at 281 G [Fig.~\ref{fig2}(c)], we estimate the magnitude of $\Pi_{\|}$ fluctuations and thereby estimate that $\Pi_{\perp}$ fluctuations are $\sim$ 1 kHz for the NV in Fig.~\ref{fig2}(c). We emphasize however that the term's nonlinearity means dephasing would be amplified beyond this estimate by static or quasistatic strain or electric fields. Second, substitutional nitrogen defects (P1 centers, $\approx$ 0.3 ppm in our implantation layer) induce magnetic fluctuations that we estimate limit DQ coherence to $\sim 100~ \mu$s \cite{Bauch2019}. Lowering the implantation dosage or coherently driving the P1 centers \cite{DeLange2012,Knowles2014,Bauch2018} would mitigate their decohering effect. Third, nitrogen implantation induces paramagnetic vacancy clusters, which we estimate also limit DQ coherence to $\sim 100 ~\mu$s \cite{FavarodeOliveira2017a}. These defects can be prevented by gentler incorporation of nitrogen such as during diamond growth \cite{Ohno2012}, or they can be removed by lattice charging during annealing \cite{FavarodeOliveira2017a} or high-temperature annealing \cite{Yamamoto2013,Tetienne2018a,Sangtawesin2018}. We remark that any remaining dephasing would be due to magnetic sources at the diamond surface that are distinct from the coherently controlled $g=2$ spins in this work.%

In conclusion, we demonstrate that the coherence time of a near-surface qubit is increased by coherently driving the surface electronic spins, without any additional materials processing or manipulation of the qubit. Using shallow NV centers as a model platform, we achieve a fivefold sensitivity enhancement by combining surface spin driving with operation in the double-quantum basis.
Future work can combine the methods presented here with existing materials processing techniques to eliminate other known sources of dephasing and push toward $T_1$-limited coherence times for shallow NV centers. The surface spin driving technique demonstrated here could also realize coherence extensions in other systems such as superconducting qubits, adatom qubits, and other near-surface qubits affected by surface spins. Further, these results also suggest other forms of driving as a promising path forward for near-surface qubits; for example, electrical driving of surface electric dipoles could suppress electric field noise that afflicts atoms and ions near surfaces \cite{McGuirk2004,Daniilidis2011,Sedlacek2018}.

\begin{acknowledgments}
We thank Tim Eichhorn, Nathalie de Leon, Hengyun Zhou, Isaac Chuang, and Viatcheslav Dobrovitski for helpful discussions. We acknowledge partial support from NSF CAREER Grant No. DMR-1352660 and partial support from the DARPA DRINQS program (agreement D18AC00014). D.B. acknowledges funding from the Microscopy Society of America and the Barry Goldwater Foundation.
\end{acknowledgments}

\bibliographystyle{apsrev4-1}
\bibliography{Surface_spin_paper_citations.bib}

\begin{thebibliography}{65}%
\makeatletter
\providecommand \@ifxundefined [1]{%
 \@ifx{#1\undefined}
}%
\providecommand \@ifnum [1]{%
 \ifnum #1\expandafter \@firstoftwo
 \else \expandafter \@secondoftwo
 \fi
}%
\providecommand \@ifx [1]{%
 \ifx #1\expandafter \@firstoftwo
 \else \expandafter \@secondoftwo
 \fi
}%
\providecommand \natexlab [1]{#1}%
\providecommand \enquote  [1]{``#1''}%
\providecommand \bibnamefont  [1]{#1}%
\providecommand \bibfnamefont [1]{#1}%
\providecommand \citenamefont [1]{#1}%
\providecommand \href@noop [0]{\@secondoftwo}%
\providecommand \href [0]{\begingroup \@sanitize@url \@href}%
\providecommand \@href[1]{\@@startlink{#1}\@@href}%
\providecommand \@@href[1]{\endgroup#1\@@endlink}%
\providecommand \@sanitize@url [0]{\catcode `\\12\catcode `\$12\catcode
  `\&12\catcode `\#12\catcode `\^12\catcode `\_12\catcode `\%12\relax}%
\providecommand \@@startlink[1]{}%
\providecommand \@@endlink[0]{}%
\providecommand \url  [0]{\begingroup\@sanitize@url \@url }%
\providecommand \@url [1]{\endgroup\@href {#1}{\urlprefix }}%
\providecommand \urlprefix  [0]{URL }%
\providecommand \Eprint [0]{\href }%
\providecommand \doibase [0]{http://dx.doi.org/}%
\providecommand \selectlanguage [0]{\@gobble}%
\providecommand \bibinfo  [0]{\@secondoftwo}%
\providecommand \bibfield  [0]{\@secondoftwo}%
\providecommand \translation [1]{[#1]}%
\providecommand \BibitemOpen [0]{}%
\providecommand \bibitemStop [0]{}%
\providecommand \bibitemNoStop [0]{.\EOS\space}%
\providecommand \EOS [0]{\spacefactor3000\relax}%
\providecommand \BibitemShut  [1]{\csname bibitem#1\endcsname}%
\let\auto@bib@innerbib\@empty
\bibitem [{\citenamefont {Wang}\ \emph {et~al.}(2015)\citenamefont {Wang},
  \citenamefont {Axline}, \citenamefont {Gao}, \citenamefont {Brecht},
  \citenamefont {Chu}, \citenamefont {Frunzio}, \citenamefont {Devoret},\ and\
  \citenamefont {Schoelkopf}}]{Wang2015}%
  \BibitemOpen
  \bibfield  {author} {\bibinfo {author} {\bibfnamefont {C.}~\bibnamefont
  {Wang}}, \bibinfo {author} {\bibfnamefont {C.}~\bibnamefont {Axline}},
  \bibinfo {author} {\bibfnamefont {Y.~Y.}\ \bibnamefont {Gao}}, \bibinfo
  {author} {\bibfnamefont {T.}~\bibnamefont {Brecht}}, \bibinfo {author}
  {\bibfnamefont {Y.}~\bibnamefont {Chu}}, \bibinfo {author} {\bibfnamefont
  {L.}~\bibnamefont {Frunzio}}, \bibinfo {author} {\bibfnamefont {M.~H.}\
  \bibnamefont {Devoret}}, \ and\ \bibinfo {author} {\bibfnamefont {R.~J.}\
  \bibnamefont {Schoelkopf}},\ }\href {\doibase 10.1063/1.4934486} {\bibfield
  {journal} {\bibinfo  {journal} {Applied Physics Letters}\ }\textbf {\bibinfo
  {volume} {107}},\ \bibinfo {pages} {162601} (\bibinfo {year}
  {2015})}\BibitemShut {NoStop}%
\bibitem [{\citenamefont {MacCabe}\ \emph {et~al.}(2019)\citenamefont
  {MacCabe}, \citenamefont {Ren}, \citenamefont {Luo}, \citenamefont {Cohen},
  \citenamefont {Zhou}, \citenamefont {Sipahigil}, \citenamefont
  {Mirhosseini},\ and\ \citenamefont {Painter}}]{MacCabe2019}%
  \BibitemOpen
  \bibfield  {author} {\bibinfo {author} {\bibfnamefont {G.~S.}\ \bibnamefont
  {MacCabe}}, \bibinfo {author} {\bibfnamefont {H.}~\bibnamefont {Ren}},
  \bibinfo {author} {\bibfnamefont {J.}~\bibnamefont {Luo}}, \bibinfo {author}
  {\bibfnamefont {J.~D.}\ \bibnamefont {Cohen}}, \bibinfo {author}
  {\bibfnamefont {H.}~\bibnamefont {Zhou}}, \bibinfo {author} {\bibfnamefont
  {A.}~\bibnamefont {Sipahigil}}, \bibinfo {author} {\bibfnamefont
  {M.}~\bibnamefont {Mirhosseini}}, \ and\ \bibinfo {author} {\bibfnamefont
  {O.}~\bibnamefont {Painter}},\ }\href {http://arxiv.org/abs/1901.04129} {\
  (\bibinfo {year} {2019})},\ \Eprint {http://arxiv.org/abs/1901.04129}
  {arXiv:1901.04129} \BibitemShut {NoStop}%
\bibitem [{\citenamefont {Baumann}\ \emph {et~al.}(2015)\citenamefont
  {Baumann}, \citenamefont {Paul}, \citenamefont {Choi}, \citenamefont {Lutz},
  \citenamefont {Ardavan},\ and\ \citenamefont {Heinrich}}]{Baumann2015}%
  \BibitemOpen
  \bibfield  {author} {\bibinfo {author} {\bibfnamefont {S.}~\bibnamefont
  {Baumann}}, \bibinfo {author} {\bibfnamefont {W.}~\bibnamefont {Paul}},
  \bibinfo {author} {\bibfnamefont {T.}~\bibnamefont {Choi}}, \bibinfo {author}
  {\bibfnamefont {C.~P.}\ \bibnamefont {Lutz}}, \bibinfo {author}
  {\bibfnamefont {A.}~\bibnamefont {Ardavan}}, \ and\ \bibinfo {author}
  {\bibfnamefont {A.~J.}\ \bibnamefont {Heinrich}},\ }\href {\doibase
  10.1126/science.aac8703} {\bibfield  {journal} {\bibinfo  {journal} {Science
  (New York, N.Y.)}\ }\textbf {\bibinfo {volume} {350}},\ \bibinfo {pages}
  {417} (\bibinfo {year} {2015})}\BibitemShut {NoStop}%
\bibitem [{\citenamefont {Lin}\ \emph {et~al.}(2004)\citenamefont {Lin},
  \citenamefont {Teper}, \citenamefont {Chin},\ and\ \citenamefont
  {Vuleti{\'{c}}}}]{Lin2004}%
  \BibitemOpen
  \bibfield  {author} {\bibinfo {author} {\bibfnamefont {Y.-j.}\ \bibnamefont
  {Lin}}, \bibinfo {author} {\bibfnamefont {I.}~\bibnamefont {Teper}}, \bibinfo
  {author} {\bibfnamefont {C.}~\bibnamefont {Chin}}, \ and\ \bibinfo {author}
  {\bibfnamefont {V.}~\bibnamefont {Vuleti{\'{c}}}},\ }\href {\doibase
  10.1103/PhysRevLett.92.050404} {\bibfield  {journal} {\bibinfo  {journal}
  {Physical Review Letters}\ }\textbf {\bibinfo {volume} {92}},\ \bibinfo
  {pages} {050404} (\bibinfo {year} {2004})}\BibitemShut {NoStop}%
\bibitem [{\citenamefont {Turchette}\ \emph {et~al.}(2000)\citenamefont
  {Turchette}, \citenamefont {Kielpinski}, \citenamefont {King}, \citenamefont
  {Leibfried}, \citenamefont {Meekhof}, \citenamefont {Myatt}, \citenamefont
  {Rowe}, \citenamefont {Sackett}, \citenamefont {Wood}, \citenamefont {Itano},
  \citenamefont {Monroe},\ and\ \citenamefont {Wineland}}]{Turchette2000}%
  \BibitemOpen
  \bibfield  {author} {\bibinfo {author} {\bibfnamefont {Q.~A.}\ \bibnamefont
  {Turchette}}, \bibinfo {author} {\bibnamefont {Kielpinski}}, \bibinfo
  {author} {\bibfnamefont {B.~E.}\ \bibnamefont {King}}, \bibinfo {author}
  {\bibfnamefont {D.}~\bibnamefont {Leibfried}}, \bibinfo {author}
  {\bibfnamefont {D.~M.}\ \bibnamefont {Meekhof}}, \bibinfo {author}
  {\bibfnamefont {C.~J.}\ \bibnamefont {Myatt}}, \bibinfo {author}
  {\bibfnamefont {M.~A.}\ \bibnamefont {Rowe}}, \bibinfo {author}
  {\bibfnamefont {C.~A.}\ \bibnamefont {Sackett}}, \bibinfo {author}
  {\bibfnamefont {C.~S.}\ \bibnamefont {Wood}}, \bibinfo {author}
  {\bibfnamefont {W.~M.}\ \bibnamefont {Itano}}, \bibinfo {author}
  {\bibfnamefont {C.}~\bibnamefont {Monroe}}, \ and\ \bibinfo {author}
  {\bibfnamefont {D.~J.}\ \bibnamefont {Wineland}},\ }\href {\doibase
  10.1103/PhysRevA.61.063418} {\bibfield  {journal} {\bibinfo  {journal}
  {Physical Review A}\ }\textbf {\bibinfo {volume} {61}},\ \bibinfo {pages}
  {063418} (\bibinfo {year} {2000})}\BibitemShut {NoStop}%
\bibitem [{\citenamefont {Myers}\ \emph {et~al.}(2014)\citenamefont {Myers},
  \citenamefont {Das}, \citenamefont {Dartiailh}, \citenamefont {Ohno},
  \citenamefont {Awschalom},\ and\ \citenamefont {{Bleszynski
  Jayich}}}]{Myers2014}%
  \BibitemOpen
  \bibfield  {author} {\bibinfo {author} {\bibfnamefont {B.}~\bibnamefont
  {Myers}}, \bibinfo {author} {\bibfnamefont {A.}~\bibnamefont {Das}}, \bibinfo
  {author} {\bibfnamefont {M.}~\bibnamefont {Dartiailh}}, \bibinfo {author}
  {\bibfnamefont {K.}~\bibnamefont {Ohno}}, \bibinfo {author} {\bibfnamefont
  {D.}~\bibnamefont {Awschalom}}, \ and\ \bibinfo {author} {\bibfnamefont
  {A.}~\bibnamefont {{Bleszynski Jayich}}},\ }\href {\doibase
  10.1103/PhysRevLett.113.027602} {\bibfield  {journal} {\bibinfo  {journal}
  {Physical Review Letters}\ }\textbf {\bibinfo {volume} {113}},\ \bibinfo
  {pages} {027602} (\bibinfo {year} {2014})}\BibitemShut {NoStop}%
\bibitem [{\citenamefont {Schenkel}\ \emph {et~al.}(2006)\citenamefont
  {Schenkel}, \citenamefont {Liddle}, \citenamefont {Persaud}, \citenamefont
  {Tyryshkin}, \citenamefont {Lyon}, \citenamefont {de~Sousa}, \citenamefont
  {Whaley}, \citenamefont {Bokor}, \citenamefont {Shangkuan},\ and\
  \citenamefont {Chakarov}}]{Schenkel2006}%
  \BibitemOpen
  \bibfield  {author} {\bibinfo {author} {\bibfnamefont {T.}~\bibnamefont
  {Schenkel}}, \bibinfo {author} {\bibfnamefont {J.~A.}\ \bibnamefont
  {Liddle}}, \bibinfo {author} {\bibfnamefont {A.}~\bibnamefont {Persaud}},
  \bibinfo {author} {\bibfnamefont {A.~M.}\ \bibnamefont {Tyryshkin}}, \bibinfo
  {author} {\bibfnamefont {S.~A.}\ \bibnamefont {Lyon}}, \bibinfo {author}
  {\bibfnamefont {R.}~\bibnamefont {de~Sousa}}, \bibinfo {author}
  {\bibfnamefont {K.~B.}\ \bibnamefont {Whaley}}, \bibinfo {author}
  {\bibfnamefont {J.}~\bibnamefont {Bokor}}, \bibinfo {author} {\bibfnamefont
  {J.}~\bibnamefont {Shangkuan}}, \ and\ \bibinfo {author} {\bibfnamefont
  {I.}~\bibnamefont {Chakarov}},\ }\href {\doibase 10.1063/1.2182068}
  {\bibfield  {journal} {\bibinfo  {journal} {Applied Physics Letters}\
  }\textbf {\bibinfo {volume} {88}},\ \bibinfo {pages} {112101} (\bibinfo
  {year} {2006})}\BibitemShut {NoStop}%
\bibitem [{\citenamefont {Sendelbach}\ \emph {et~al.}(2008)\citenamefont
  {Sendelbach}, \citenamefont {Hover}, \citenamefont {Kittel}, \citenamefont
  {M{\"{u}}ck}, \citenamefont {Martinis},\ and\ \citenamefont
  {McDermott}}]{Sendelbach2008}%
  \BibitemOpen
  \bibfield  {author} {\bibinfo {author} {\bibfnamefont {S.}~\bibnamefont
  {Sendelbach}}, \bibinfo {author} {\bibfnamefont {D.}~\bibnamefont {Hover}},
  \bibinfo {author} {\bibfnamefont {A.}~\bibnamefont {Kittel}}, \bibinfo
  {author} {\bibfnamefont {M.}~\bibnamefont {M{\"{u}}ck}}, \bibinfo {author}
  {\bibfnamefont {J.~M.}\ \bibnamefont {Martinis}}, \ and\ \bibinfo {author}
  {\bibfnamefont {R.}~\bibnamefont {McDermott}},\ }\href {\doibase
  10.1103/PhysRevLett.100.227006} {\bibfield  {journal} {\bibinfo  {journal}
  {Physical Review Letters}\ }\textbf {\bibinfo {volume} {100}},\ \bibinfo
  {pages} {227006} (\bibinfo {year} {2008})}\BibitemShut {NoStop}%
\bibitem [{\citenamefont {Bluhm}\ \emph {et~al.}(2009)\citenamefont {Bluhm},
  \citenamefont {Bert}, \citenamefont {Koshnick}, \citenamefont {Huber},\ and\
  \citenamefont {Moler}}]{Bluhm2009}%
  \BibitemOpen
  \bibfield  {author} {\bibinfo {author} {\bibfnamefont {H.}~\bibnamefont
  {Bluhm}}, \bibinfo {author} {\bibfnamefont {J.~A.}\ \bibnamefont {Bert}},
  \bibinfo {author} {\bibfnamefont {N.~C.}\ \bibnamefont {Koshnick}}, \bibinfo
  {author} {\bibfnamefont {M.~E.}\ \bibnamefont {Huber}}, \ and\ \bibinfo
  {author} {\bibfnamefont {K.~A.}\ \bibnamefont {Moler}},\ }\href {\doibase
  10.1103/PhysRevLett.103.026805} {\bibfield  {journal} {\bibinfo  {journal}
  {Physical Review Letters}\ }\textbf {\bibinfo {volume} {103}},\ \bibinfo
  {pages} {026805} (\bibinfo {year} {2009})}\BibitemShut {NoStop}%
\bibitem [{\citenamefont {Ofori-Okai}\ \emph {et~al.}(2012)\citenamefont
  {Ofori-Okai}, \citenamefont {Pezzagna}, \citenamefont {Chang}, \citenamefont
  {Loretz}, \citenamefont {Schirhagl}, \citenamefont {Tao}, \citenamefont
  {Moores}, \citenamefont {Groot-Berning}, \citenamefont {Meijer},\ and\
  \citenamefont {Degen}}]{Ofori-Okai2012b}%
  \BibitemOpen
  \bibfield  {author} {\bibinfo {author} {\bibfnamefont {B.~K.}\ \bibnamefont
  {Ofori-Okai}}, \bibinfo {author} {\bibfnamefont {S.}~\bibnamefont
  {Pezzagna}}, \bibinfo {author} {\bibfnamefont {K.}~\bibnamefont {Chang}},
  \bibinfo {author} {\bibfnamefont {M.}~\bibnamefont {Loretz}}, \bibinfo
  {author} {\bibfnamefont {R.}~\bibnamefont {Schirhagl}}, \bibinfo {author}
  {\bibfnamefont {Y.}~\bibnamefont {Tao}}, \bibinfo {author} {\bibfnamefont
  {B.~A.}\ \bibnamefont {Moores}}, \bibinfo {author} {\bibfnamefont
  {K.}~\bibnamefont {Groot-Berning}}, \bibinfo {author} {\bibfnamefont
  {J.}~\bibnamefont {Meijer}}, \ and\ \bibinfo {author} {\bibfnamefont {C.~L.}\
  \bibnamefont {Degen}},\ }\href {\doibase 10.1103/PhysRevB.86.081406}
  {\bibfield  {journal} {\bibinfo  {journal} {Physical Review B}\ }\textbf
  {\bibinfo {volume} {86}},\ \bibinfo {pages} {081406} (\bibinfo {year}
  {2012})}\BibitemShut {NoStop}%
\bibitem [{\citenamefont {Rosskopf}\ \emph {et~al.}(2014)\citenamefont
  {Rosskopf}, \citenamefont {Dussaux}, \citenamefont {Ohashi}, \citenamefont
  {Loretz}, \citenamefont {Schirhagl}, \citenamefont {Watanabe}, \citenamefont
  {Shikata}, \citenamefont {Itoh},\ and\ \citenamefont {Degen}}]{Rosskopf2014}%
  \BibitemOpen
  \bibfield  {author} {\bibinfo {author} {\bibfnamefont {T.}~\bibnamefont
  {Rosskopf}}, \bibinfo {author} {\bibfnamefont {A.}~\bibnamefont {Dussaux}},
  \bibinfo {author} {\bibfnamefont {K.}~\bibnamefont {Ohashi}}, \bibinfo
  {author} {\bibfnamefont {M.}~\bibnamefont {Loretz}}, \bibinfo {author}
  {\bibfnamefont {R.}~\bibnamefont {Schirhagl}}, \bibinfo {author}
  {\bibfnamefont {H.}~\bibnamefont {Watanabe}}, \bibinfo {author}
  {\bibfnamefont {S.}~\bibnamefont {Shikata}}, \bibinfo {author} {\bibfnamefont
  {K.~M.}\ \bibnamefont {Itoh}}, \ and\ \bibinfo {author} {\bibfnamefont
  {C.~L.}\ \bibnamefont {Degen}},\ }\href {\doibase
  10.1103/PhysRevLett.112.147602} {\bibfield  {journal} {\bibinfo  {journal}
  {Physical Review Letters}\ }\textbf {\bibinfo {volume} {112}},\ \bibinfo
  {pages} {147602} (\bibinfo {year} {2014})}\BibitemShut {NoStop}%
\bibitem [{\citenamefont {Romach}\ \emph {et~al.}(2015)\citenamefont {Romach},
  \citenamefont {M{\"{u}}ller}, \citenamefont {Unden}, \citenamefont {Rogers},
  \citenamefont {Isoda}, \citenamefont {Itoh}, \citenamefont {Markham},
  \citenamefont {Stacey}, \citenamefont {Meijer}, \citenamefont {Pezzagna},
  \citenamefont {Naydenov}, \citenamefont {McGuinness}, \citenamefont
  {Bar-Gill},\ and\ \citenamefont {Jelezko}}]{Romach2015}%
  \BibitemOpen
  \bibfield  {author} {\bibinfo {author} {\bibfnamefont {Y.}~\bibnamefont
  {Romach}}, \bibinfo {author} {\bibfnamefont {C.}~\bibnamefont
  {M{\"{u}}ller}}, \bibinfo {author} {\bibfnamefont {T.}~\bibnamefont {Unden}},
  \bibinfo {author} {\bibfnamefont {L.}~\bibnamefont {Rogers}}, \bibinfo
  {author} {\bibfnamefont {T.}~\bibnamefont {Isoda}}, \bibinfo {author}
  {\bibfnamefont {K.}~\bibnamefont {Itoh}}, \bibinfo {author} {\bibfnamefont
  {M.}~\bibnamefont {Markham}}, \bibinfo {author} {\bibfnamefont
  {A.}~\bibnamefont {Stacey}}, \bibinfo {author} {\bibfnamefont
  {J.}~\bibnamefont {Meijer}}, \bibinfo {author} {\bibfnamefont
  {S.}~\bibnamefont {Pezzagna}}, \bibinfo {author} {\bibfnamefont
  {B.}~\bibnamefont {Naydenov}}, \bibinfo {author} {\bibfnamefont
  {L.}~\bibnamefont {McGuinness}}, \bibinfo {author} {\bibfnamefont
  {N.}~\bibnamefont {Bar-Gill}}, \ and\ \bibinfo {author} {\bibfnamefont
  {F.}~\bibnamefont {Jelezko}},\ }\href {\doibase
  10.1103/PhysRevLett.114.017601} {\bibfield  {journal} {\bibinfo  {journal}
  {Physical Review Letters}\ }\textbf {\bibinfo {volume} {114}},\ \bibinfo
  {pages} {017601} (\bibinfo {year} {2015})}\BibitemShut {NoStop}%
\bibitem [{\citenamefont {de~Graaf}\ \emph {et~al.}(2017)\citenamefont
  {de~Graaf}, \citenamefont {Adamyan}, \citenamefont {Lindstr{\"{o}}m},
  \citenamefont {Erts}, \citenamefont {Kubatkin}, \citenamefont {Tzalenchuk},\
  and\ \citenamefont {Danilov}}]{DeGraaf2017}%
  \BibitemOpen
  \bibfield  {author} {\bibinfo {author} {\bibfnamefont {S.~E.}\ \bibnamefont
  {de~Graaf}}, \bibinfo {author} {\bibfnamefont {A.~A.}\ \bibnamefont
  {Adamyan}}, \bibinfo {author} {\bibfnamefont {T.}~\bibnamefont
  {Lindstr{\"{o}}m}}, \bibinfo {author} {\bibfnamefont {D.}~\bibnamefont
  {Erts}}, \bibinfo {author} {\bibfnamefont {S.~E.}\ \bibnamefont {Kubatkin}},
  \bibinfo {author} {\bibfnamefont {A.~Y.}\ \bibnamefont {Tzalenchuk}}, \ and\
  \bibinfo {author} {\bibfnamefont {A.~V.}\ \bibnamefont {Danilov}},\ }\href
  {\doibase 10.1103/PhysRevLett.118.057703} {\bibfield  {journal} {\bibinfo
  {journal} {Physical Review Letters}\ }\textbf {\bibinfo {volume} {118}},\
  \bibinfo {pages} {057703} (\bibinfo {year} {2017})}\BibitemShut {NoStop}%
\bibitem [{\citenamefont {Doherty}\ \emph {et~al.}(2013)\citenamefont
  {Doherty}, \citenamefont {Manson}, \citenamefont {Delaney}, \citenamefont
  {Jelezko}, \citenamefont {Wrachtrup},\ and\ \citenamefont
  {Hollenberg}}]{Doherty2013a}%
  \BibitemOpen
  \bibfield  {author} {\bibinfo {author} {\bibfnamefont {M.~W.}\ \bibnamefont
  {Doherty}}, \bibinfo {author} {\bibfnamefont {N.~B.}\ \bibnamefont {Manson}},
  \bibinfo {author} {\bibfnamefont {P.}~\bibnamefont {Delaney}}, \bibinfo
  {author} {\bibfnamefont {F.}~\bibnamefont {Jelezko}}, \bibinfo {author}
  {\bibfnamefont {J.}~\bibnamefont {Wrachtrup}}, \ and\ \bibinfo {author}
  {\bibfnamefont {L.~C.}\ \bibnamefont {Hollenberg}},\ }\href {\doibase
  10.1016/J.PHYSREP.2013.02.001} {\bibfield  {journal} {\bibinfo  {journal}
  {Physics Reports}\ }\textbf {\bibinfo {volume} {528}},\ \bibinfo {pages} {1}
  (\bibinfo {year} {2013})}\BibitemShut {NoStop}%
\bibitem [{\citenamefont {Staudacher}\ \emph {et~al.}(2013)\citenamefont
  {Staudacher}, \citenamefont {Shi}, \citenamefont {Pezzagna}, \citenamefont
  {Meijer}, \citenamefont {Du}, \citenamefont {Meriles}, \citenamefont
  {Reinhard},\ and\ \citenamefont {Wrachtrup}}]{Staudacher2013a}%
  \BibitemOpen
  \bibfield  {author} {\bibinfo {author} {\bibfnamefont {T.}~\bibnamefont
  {Staudacher}}, \bibinfo {author} {\bibfnamefont {F.}~\bibnamefont {Shi}},
  \bibinfo {author} {\bibfnamefont {S.}~\bibnamefont {Pezzagna}}, \bibinfo
  {author} {\bibfnamefont {J.}~\bibnamefont {Meijer}}, \bibinfo {author}
  {\bibfnamefont {J.}~\bibnamefont {Du}}, \bibinfo {author} {\bibfnamefont
  {C.~A.}\ \bibnamefont {Meriles}}, \bibinfo {author} {\bibfnamefont
  {F.}~\bibnamefont {Reinhard}}, \ and\ \bibinfo {author} {\bibfnamefont
  {J.}~\bibnamefont {Wrachtrup}},\ }\href {\doibase 10.1126/science.1231675}
  {\bibfield  {journal} {\bibinfo  {journal} {Science (New York, N.Y.)}\
  }\textbf {\bibinfo {volume} {339}},\ \bibinfo {pages} {561} (\bibinfo {year}
  {2013})}\BibitemShut {NoStop}%
\bibitem [{\citenamefont {Gross}\ \emph {et~al.}(2017)\citenamefont {Gross},
  \citenamefont {Akhtar}, \citenamefont {Garcia}, \citenamefont
  {Mart{\'{i}}nez}, \citenamefont {Chouaieb}, \citenamefont {Garcia},
  \citenamefont {Carr{\'{e}}t{\'{e}}ro}, \citenamefont
  {Barth{\'{e}}l{\'{e}}my}, \citenamefont {Appel}, \citenamefont {Maletinsky},
  \citenamefont {Kim}, \citenamefont {Chauleau}, \citenamefont {Jaouen},
  \citenamefont {Viret}, \citenamefont {Bibes}, \citenamefont {Fusil},\ and\
  \citenamefont {Jacques}}]{Gross2017}%
  \BibitemOpen
  \bibfield  {author} {\bibinfo {author} {\bibfnamefont {I.}~\bibnamefont
  {Gross}}, \bibinfo {author} {\bibfnamefont {W.}~\bibnamefont {Akhtar}},
  \bibinfo {author} {\bibfnamefont {V.}~\bibnamefont {Garcia}}, \bibinfo
  {author} {\bibfnamefont {L.~J.}\ \bibnamefont {Mart{\'{i}}nez}}, \bibinfo
  {author} {\bibfnamefont {S.}~\bibnamefont {Chouaieb}}, \bibinfo {author}
  {\bibfnamefont {K.}~\bibnamefont {Garcia}}, \bibinfo {author} {\bibfnamefont
  {C.}~\bibnamefont {Carr{\'{e}}t{\'{e}}ro}}, \bibinfo {author} {\bibfnamefont
  {A.}~\bibnamefont {Barth{\'{e}}l{\'{e}}my}}, \bibinfo {author} {\bibfnamefont
  {P.}~\bibnamefont {Appel}}, \bibinfo {author} {\bibfnamefont
  {P.}~\bibnamefont {Maletinsky}}, \bibinfo {author} {\bibfnamefont {J.-V.}\
  \bibnamefont {Kim}}, \bibinfo {author} {\bibfnamefont {J.~Y.}\ \bibnamefont
  {Chauleau}}, \bibinfo {author} {\bibfnamefont {N.}~\bibnamefont {Jaouen}},
  \bibinfo {author} {\bibfnamefont {M.}~\bibnamefont {Viret}}, \bibinfo
  {author} {\bibfnamefont {M.}~\bibnamefont {Bibes}}, \bibinfo {author}
  {\bibfnamefont {S.}~\bibnamefont {Fusil}}, \ and\ \bibinfo {author}
  {\bibfnamefont {V.}~\bibnamefont {Jacques}},\ }\href {\doibase
  10.1038/nature23656} {\bibfield  {journal} {\bibinfo  {journal} {Nature}\
  }\textbf {\bibinfo {volume} {549}},\ \bibinfo {pages} {252} (\bibinfo {year}
  {2017})}\BibitemShut {NoStop}%
\bibitem [{\citenamefont {Casola}\ \emph {et~al.}(2018)\citenamefont {Casola},
  \citenamefont {van~der Sar},\ and\ \citenamefont {Yacoby}}]{Casola2018}%
  \BibitemOpen
  \bibfield  {author} {\bibinfo {author} {\bibfnamefont {F.}~\bibnamefont
  {Casola}}, \bibinfo {author} {\bibfnamefont {T.}~\bibnamefont {van~der Sar}},
  \ and\ \bibinfo {author} {\bibfnamefont {A.}~\bibnamefont {Yacoby}},\ }\href
  {\doibase 10.1038/natrevmats.2017.88} {\bibfield  {journal} {\bibinfo
  {journal} {Nature Reviews Materials}\ }\textbf {\bibinfo {volume} {3}},\
  \bibinfo {pages} {17088} (\bibinfo {year} {2018})}\BibitemShut {NoStop}%
\bibitem [{\citenamefont {Ariyaratne}\ \emph {et~al.}(2018)\citenamefont
  {Ariyaratne}, \citenamefont {Bluvstein}, \citenamefont {Myers},\ and\
  \citenamefont {Jayich}}]{Ariyaratne2018b}%
  \BibitemOpen
  \bibfield  {author} {\bibinfo {author} {\bibfnamefont {A.}~\bibnamefont
  {Ariyaratne}}, \bibinfo {author} {\bibfnamefont {D.}~\bibnamefont
  {Bluvstein}}, \bibinfo {author} {\bibfnamefont {B.~A.}\ \bibnamefont
  {Myers}}, \ and\ \bibinfo {author} {\bibfnamefont {A.~C.~B.}\ \bibnamefont
  {Jayich}},\ }\href {\doibase 10.1038/s41467-018-04798-1} {\bibfield
  {journal} {\bibinfo  {journal} {Nature Communications}\ }\textbf {\bibinfo
  {volume} {9}},\ \bibinfo {pages} {2406} (\bibinfo {year} {2018})}\BibitemShut
  {NoStop}%
\bibitem [{\citenamefont {Faraon}\ \emph {et~al.}(2011)\citenamefont {Faraon},
  \citenamefont {Barclay}, \citenamefont {Santori}, \citenamefont {Fu},\ and\
  \citenamefont {Beausoleil}}]{Faraon2011a}%
  \BibitemOpen
  \bibfield  {author} {\bibinfo {author} {\bibfnamefont {A.}~\bibnamefont
  {Faraon}}, \bibinfo {author} {\bibfnamefont {P.~E.}\ \bibnamefont {Barclay}},
  \bibinfo {author} {\bibfnamefont {C.}~\bibnamefont {Santori}}, \bibinfo
  {author} {\bibfnamefont {K.-M.~C.}\ \bibnamefont {Fu}}, \ and\ \bibinfo
  {author} {\bibfnamefont {R.~G.}\ \bibnamefont {Beausoleil}},\ }\href
  {\doibase 10.1038/nphoton.2011.52} {\bibfield  {journal} {\bibinfo  {journal}
  {Nature Photonics}\ }\textbf {\bibinfo {volume} {5}},\ \bibinfo {pages} {301}
  (\bibinfo {year} {2011})}\BibitemShut {NoStop}%
\bibitem [{\citenamefont {Cai}\ \emph {et~al.}(2013)\citenamefont {Cai},
  \citenamefont {Retzker}, \citenamefont {Jelezko},\ and\ \citenamefont
  {Plenio}}]{Cai2013a}%
  \BibitemOpen
  \bibfield  {author} {\bibinfo {author} {\bibfnamefont {J.}~\bibnamefont
  {Cai}}, \bibinfo {author} {\bibfnamefont {A.}~\bibnamefont {Retzker}},
  \bibinfo {author} {\bibfnamefont {F.}~\bibnamefont {Jelezko}}, \ and\
  \bibinfo {author} {\bibfnamefont {M.~B.}\ \bibnamefont {Plenio}},\ }\href
  {\doibase 10.1038/nphys2519} {\bibfield  {journal} {\bibinfo  {journal}
  {Nature Physics}\ }\textbf {\bibinfo {volume} {9}},\ \bibinfo {pages} {168}
  (\bibinfo {year} {2013})}\BibitemShut {NoStop}%
\bibitem [{\citenamefont {Golter}\ \emph {et~al.}(2016)\citenamefont {Golter},
  \citenamefont {Oo}, \citenamefont {Amezcua}, \citenamefont {Stewart},\ and\
  \citenamefont {Wang}}]{Golter2016}%
  \BibitemOpen
  \bibfield  {author} {\bibinfo {author} {\bibfnamefont {D.~A.}\ \bibnamefont
  {Golter}}, \bibinfo {author} {\bibfnamefont {T.}~\bibnamefont {Oo}}, \bibinfo
  {author} {\bibfnamefont {M.}~\bibnamefont {Amezcua}}, \bibinfo {author}
  {\bibfnamefont {K.~A.}\ \bibnamefont {Stewart}}, \ and\ \bibinfo {author}
  {\bibfnamefont {H.}~\bibnamefont {Wang}},\ }\href {\doibase
  10.1103/PhysRevLett.116.143602} {\bibfield  {journal} {\bibinfo  {journal}
  {Physical Review Letters}\ }\textbf {\bibinfo {volume} {116}},\ \bibinfo
  {pages} {143602} (\bibinfo {year} {2016})}\BibitemShut {NoStop}%
\bibitem [{\citenamefont {Lovchinsky}\ \emph {et~al.}(2016)\citenamefont
  {Lovchinsky}, \citenamefont {Sushkov}, \citenamefont {Urbach}, \citenamefont
  {de~Leon}, \citenamefont {Choi}, \citenamefont {{De Greve}}, \citenamefont
  {Evans}, \citenamefont {Gertner}, \citenamefont {Bersin}, \citenamefont
  {M{\"{u}}ller}, \citenamefont {McGuinness}, \citenamefont {Jelezko},
  \citenamefont {Walsworth}, \citenamefont {Park},\ and\ \citenamefont
  {Lukin}}]{Lovchinsky2016a}%
  \BibitemOpen
  \bibfield  {author} {\bibinfo {author} {\bibfnamefont {I.}~\bibnamefont
  {Lovchinsky}}, \bibinfo {author} {\bibfnamefont {A.~O.}\ \bibnamefont
  {Sushkov}}, \bibinfo {author} {\bibfnamefont {E.}~\bibnamefont {Urbach}},
  \bibinfo {author} {\bibfnamefont {N.~P.}\ \bibnamefont {de~Leon}}, \bibinfo
  {author} {\bibfnamefont {S.}~\bibnamefont {Choi}}, \bibinfo {author}
  {\bibfnamefont {K.}~\bibnamefont {{De Greve}}}, \bibinfo {author}
  {\bibfnamefont {R.}~\bibnamefont {Evans}}, \bibinfo {author} {\bibfnamefont
  {R.}~\bibnamefont {Gertner}}, \bibinfo {author} {\bibfnamefont
  {E.}~\bibnamefont {Bersin}}, \bibinfo {author} {\bibfnamefont
  {C.}~\bibnamefont {M{\"{u}}ller}}, \bibinfo {author} {\bibfnamefont
  {L.}~\bibnamefont {McGuinness}}, \bibinfo {author} {\bibfnamefont
  {F.}~\bibnamefont {Jelezko}}, \bibinfo {author} {\bibfnamefont {R.~L.}\
  \bibnamefont {Walsworth}}, \bibinfo {author} {\bibfnamefont {H.}~\bibnamefont
  {Park}}, \ and\ \bibinfo {author} {\bibfnamefont {M.~D.}\ \bibnamefont
  {Lukin}},\ }\href {\doibase 10.1126/science.aad8022} {\bibfield  {journal}
  {\bibinfo  {journal} {Science (New York, N.Y.)}\ }\textbf {\bibinfo {volume}
  {351}},\ \bibinfo {pages} {836} (\bibinfo {year} {2016})}\BibitemShut
  {NoStop}%
\bibitem [{\citenamefont {Degen}\ \emph {et~al.}(2017)\citenamefont {Degen},
  \citenamefont {Reinhard},\ and\ \citenamefont {Cappellaro}}]{Degen2017}%
  \BibitemOpen
  \bibfield  {author} {\bibinfo {author} {\bibfnamefont {C.~L.}\ \bibnamefont
  {Degen}}, \bibinfo {author} {\bibfnamefont {F.}~\bibnamefont {Reinhard}}, \
  and\ \bibinfo {author} {\bibfnamefont {P.}~\bibnamefont {Cappellaro}},\
  }\href {\doibase 10.1103/RevModPhys.89.035002} {\bibfield  {journal}
  {\bibinfo  {journal} {Reviews of Modern Physics}\ }\textbf {\bibinfo {volume}
  {89}},\ \bibinfo {pages} {035002} (\bibinfo {year} {2017})}\BibitemShut
  {NoStop}%
\bibitem [{\citenamefont {Taylor}\ \emph {et~al.}(2008)\citenamefont {Taylor},
  \citenamefont {Cappellaro}, \citenamefont {Childress}, \citenamefont {Jiang},
  \citenamefont {Budker}, \citenamefont {Hemmer}, \citenamefont {Yacoby},
  \citenamefont {Walsworth},\ and\ \citenamefont {Lukin}}]{Taylor2008b}%
  \BibitemOpen
  \bibfield  {author} {\bibinfo {author} {\bibfnamefont {J.~M.}\ \bibnamefont
  {Taylor}}, \bibinfo {author} {\bibfnamefont {P.}~\bibnamefont {Cappellaro}},
  \bibinfo {author} {\bibfnamefont {L.}~\bibnamefont {Childress}}, \bibinfo
  {author} {\bibfnamefont {L.}~\bibnamefont {Jiang}}, \bibinfo {author}
  {\bibfnamefont {D.}~\bibnamefont {Budker}}, \bibinfo {author} {\bibfnamefont
  {P.~R.}\ \bibnamefont {Hemmer}}, \bibinfo {author} {\bibfnamefont
  {A.}~\bibnamefont {Yacoby}}, \bibinfo {author} {\bibfnamefont
  {R.}~\bibnamefont {Walsworth}}, \ and\ \bibinfo {author} {\bibfnamefont
  {M.~D.}\ \bibnamefont {Lukin}},\ }\href {\doibase 10.1038/nphys1075}
  {\bibfield  {journal} {\bibinfo  {journal} {Nature Physics}\ }\textbf
  {\bibinfo {volume} {4}},\ \bibinfo {pages} {810} (\bibinfo {year}
  {2008})}\BibitemShut {NoStop}%
\bibitem [{\citenamefont {Pelliccione}\ \emph {et~al.}(2016)\citenamefont
  {Pelliccione}, \citenamefont {Jenkins}, \citenamefont {Ovartchaiyapong},
  \citenamefont {Reetz}, \citenamefont {Emmanouilidou}, \citenamefont {Ni},\
  and\ \citenamefont {{Bleszynski Jayich}}}]{Pelliccione2016b}%
  \BibitemOpen
  \bibfield  {author} {\bibinfo {author} {\bibfnamefont {M.}~\bibnamefont
  {Pelliccione}}, \bibinfo {author} {\bibfnamefont {A.}~\bibnamefont
  {Jenkins}}, \bibinfo {author} {\bibfnamefont {P.}~\bibnamefont
  {Ovartchaiyapong}}, \bibinfo {author} {\bibfnamefont {C.}~\bibnamefont
  {Reetz}}, \bibinfo {author} {\bibfnamefont {E.}~\bibnamefont
  {Emmanouilidou}}, \bibinfo {author} {\bibfnamefont {N.}~\bibnamefont {Ni}}, \
  and\ \bibinfo {author} {\bibfnamefont {A.~C.}\ \bibnamefont {{Bleszynski
  Jayich}}},\ }\href {\doibase 10.1038/nnano.2016.68} {\bibfield  {journal}
  {\bibinfo  {journal} {Nature Nanotechnology}\ }\textbf {\bibinfo {volume}
  {11}},\ \bibinfo {pages} {700} (\bibinfo {year} {2016})}\BibitemShut
  {NoStop}%
\bibitem [{\citenamefont {Kim}\ \emph {et~al.}(2015)\citenamefont {Kim},
  \citenamefont {Mamin}, \citenamefont {Sherwood}, \citenamefont {Ohno},
  \citenamefont {Awschalom},\ and\ \citenamefont {Rugar}}]{Kim2015}%
  \BibitemOpen
  \bibfield  {author} {\bibinfo {author} {\bibfnamefont {M.}~\bibnamefont
  {Kim}}, \bibinfo {author} {\bibfnamefont {H.~J.}\ \bibnamefont {Mamin}},
  \bibinfo {author} {\bibfnamefont {M.~H.}\ \bibnamefont {Sherwood}}, \bibinfo
  {author} {\bibfnamefont {K.}~\bibnamefont {Ohno}}, \bibinfo {author}
  {\bibfnamefont {D.~D.}\ \bibnamefont {Awschalom}}, \ and\ \bibinfo {author}
  {\bibfnamefont {D.}~\bibnamefont {Rugar}},\ }\href {\doibase
  10.1103/PhysRevLett.115.087602} {\bibfield  {journal} {\bibinfo  {journal}
  {Physical Review Letters}\ }\textbf {\bibinfo {volume} {115}},\ \bibinfo
  {pages} {087602} (\bibinfo {year} {2015})}\BibitemShut {NoStop}%
\bibitem [{\citenamefont {Myers}\ \emph {et~al.}(2017)\citenamefont {Myers},
  \citenamefont {Ariyaratne},\ and\ \citenamefont {Jayich}}]{Myers2017b}%
  \BibitemOpen
  \bibfield  {author} {\bibinfo {author} {\bibfnamefont {B.}~\bibnamefont
  {Myers}}, \bibinfo {author} {\bibfnamefont {A.}~\bibnamefont {Ariyaratne}}, \
  and\ \bibinfo {author} {\bibfnamefont {A.~B.}\ \bibnamefont {Jayich}},\
  }\href {\doibase 10.1103/PhysRevLett.118.197201} {\bibfield  {journal}
  {\bibinfo  {journal} {Physical Review Letters}\ }\textbf {\bibinfo {volume}
  {118}},\ \bibinfo {pages} {197201} (\bibinfo {year} {2017})}\BibitemShut
  {NoStop}%
\bibitem [{\citenamefont {Tisler}\ \emph {et~al.}(2009)\citenamefont {Tisler},
  \citenamefont {Balasubramanian}, \citenamefont {Naydenov}, \citenamefont
  {Kolesov}, \citenamefont {Grotz}, \citenamefont {Reuter}, \citenamefont
  {Boudou}, \citenamefont {Curmi}, \citenamefont {Sennour}, \citenamefont
  {Thorel}, \citenamefont {B{\"{o}}rsch}, \citenamefont {Aulenbacher},
  \citenamefont {Erdmann}, \citenamefont {Hemmer}, \citenamefont {Jelezko},\
  and\ \citenamefont {Wrachtrup}}]{Tisler2009}%
  \BibitemOpen
  \bibfield  {author} {\bibinfo {author} {\bibfnamefont {J.}~\bibnamefont
  {Tisler}}, \bibinfo {author} {\bibfnamefont {G.}~\bibnamefont
  {Balasubramanian}}, \bibinfo {author} {\bibfnamefont {B.}~\bibnamefont
  {Naydenov}}, \bibinfo {author} {\bibfnamefont {R.}~\bibnamefont {Kolesov}},
  \bibinfo {author} {\bibfnamefont {B.}~\bibnamefont {Grotz}}, \bibinfo
  {author} {\bibfnamefont {R.}~\bibnamefont {Reuter}}, \bibinfo {author}
  {\bibfnamefont {J.-P.}\ \bibnamefont {Boudou}}, \bibinfo {author}
  {\bibfnamefont {P.~A.}\ \bibnamefont {Curmi}}, \bibinfo {author}
  {\bibfnamefont {M.}~\bibnamefont {Sennour}}, \bibinfo {author} {\bibfnamefont
  {A.}~\bibnamefont {Thorel}}, \bibinfo {author} {\bibfnamefont
  {M.}~\bibnamefont {B{\"{o}}rsch}}, \bibinfo {author} {\bibfnamefont
  {K.}~\bibnamefont {Aulenbacher}}, \bibinfo {author} {\bibfnamefont
  {R.}~\bibnamefont {Erdmann}}, \bibinfo {author} {\bibfnamefont {P.~R.}\
  \bibnamefont {Hemmer}}, \bibinfo {author} {\bibfnamefont {F.}~\bibnamefont
  {Jelezko}}, \ and\ \bibinfo {author} {\bibfnamefont {J.}~\bibnamefont
  {Wrachtrup}},\ }\href {\doibase 10.1021/nn9003617} {\bibfield  {journal}
  {\bibinfo  {journal} {ACS Nano}\ }\textbf {\bibinfo {volume} {3}},\ \bibinfo
  {pages} {1959} (\bibinfo {year} {2009})}\BibitemShut {NoStop}%
\bibitem [{\citenamefont {Hite}\ \emph {et~al.}(2013)\citenamefont {Hite},
  \citenamefont {Colombe}, \citenamefont {Wilson}, \citenamefont {Allcock},
  \citenamefont {Leibfried}, \citenamefont {Wineland},\ and\ \citenamefont
  {Pappas}}]{Hite2013}%
  \BibitemOpen
  \bibfield  {author} {\bibinfo {author} {\bibfnamefont {D.}~\bibnamefont
  {Hite}}, \bibinfo {author} {\bibfnamefont {Y.}~\bibnamefont {Colombe}},
  \bibinfo {author} {\bibfnamefont {A.}~\bibnamefont {Wilson}}, \bibinfo
  {author} {\bibfnamefont {D.}~\bibnamefont {Allcock}}, \bibinfo {author}
  {\bibfnamefont {D.}~\bibnamefont {Leibfried}}, \bibinfo {author}
  {\bibfnamefont {D.}~\bibnamefont {Wineland}}, \ and\ \bibinfo {author}
  {\bibfnamefont {D.}~\bibnamefont {Pappas}},\ }\href {\doibase
  10.1557/mrs.2013.207} {\bibfield  {journal} {\bibinfo  {journal} {MRS
  Bulletin}\ }\textbf {\bibinfo {volume} {38}},\ \bibinfo {pages} {826}
  (\bibinfo {year} {2013})}\BibitemShut {NoStop}%
\bibitem [{\citenamefont {Quintana}\ \emph {et~al.}(2014)\citenamefont
  {Quintana}, \citenamefont {Megrant}, \citenamefont {Chen}, \citenamefont
  {Dunsworth}, \citenamefont {Chiaro}, \citenamefont {Barends}, \citenamefont
  {Campbell}, \citenamefont {Chen}, \citenamefont {Hoi}, \citenamefont
  {Jeffrey}, \citenamefont {Kelly}, \citenamefont {Mutus}, \citenamefont
  {O'Malley}, \citenamefont {Neill}, \citenamefont {Roushan}, \citenamefont
  {Sank}, \citenamefont {Vainsencher}, \citenamefont {Wenner}, \citenamefont
  {White}, \citenamefont {Cleland},\ and\ \citenamefont
  {Martinis}}]{Quintana2014}%
  \BibitemOpen
  \bibfield  {author} {\bibinfo {author} {\bibfnamefont {C.~M.}\ \bibnamefont
  {Quintana}}, \bibinfo {author} {\bibfnamefont {A.}~\bibnamefont {Megrant}},
  \bibinfo {author} {\bibfnamefont {Z.}~\bibnamefont {Chen}}, \bibinfo {author}
  {\bibfnamefont {A.}~\bibnamefont {Dunsworth}}, \bibinfo {author}
  {\bibfnamefont {B.}~\bibnamefont {Chiaro}}, \bibinfo {author} {\bibfnamefont
  {R.}~\bibnamefont {Barends}}, \bibinfo {author} {\bibfnamefont
  {B.}~\bibnamefont {Campbell}}, \bibinfo {author} {\bibfnamefont
  {Y.}~\bibnamefont {Chen}}, \bibinfo {author} {\bibfnamefont {I.-C.}\
  \bibnamefont {Hoi}}, \bibinfo {author} {\bibfnamefont {E.}~\bibnamefont
  {Jeffrey}}, \bibinfo {author} {\bibfnamefont {J.}~\bibnamefont {Kelly}},
  \bibinfo {author} {\bibfnamefont {J.~Y.}\ \bibnamefont {Mutus}}, \bibinfo
  {author} {\bibfnamefont {P.~J.~J.}\ \bibnamefont {O'Malley}}, \bibinfo
  {author} {\bibfnamefont {C.}~\bibnamefont {Neill}}, \bibinfo {author}
  {\bibfnamefont {P.}~\bibnamefont {Roushan}}, \bibinfo {author} {\bibfnamefont
  {D.}~\bibnamefont {Sank}}, \bibinfo {author} {\bibfnamefont {A.}~\bibnamefont
  {Vainsencher}}, \bibinfo {author} {\bibfnamefont {J.}~\bibnamefont {Wenner}},
  \bibinfo {author} {\bibfnamefont {T.~C.}\ \bibnamefont {White}}, \bibinfo
  {author} {\bibfnamefont {A.~N.}\ \bibnamefont {Cleland}}, \ and\ \bibinfo
  {author} {\bibfnamefont {J.~M.}\ \bibnamefont {Martinis}},\ }\href {\doibase
  10.1063/1.4893297} {\bibfield  {journal} {\bibinfo  {journal} {Applied
  Physics Letters}\ }\textbf {\bibinfo {volume} {105}},\ \bibinfo {pages}
  {062601} (\bibinfo {year} {2014})}\BibitemShut {NoStop}%
\bibitem [{\citenamefont {Kumar}\ \emph {et~al.}(2016)\citenamefont {Kumar},
  \citenamefont {Sendelbach}, \citenamefont {Beck}, \citenamefont {Freeland},
  \citenamefont {Wang}, \citenamefont {Wang}, \citenamefont {Yu}, \citenamefont
  {Wu}, \citenamefont {Pappas},\ and\ \citenamefont {McDermott}}]{Kumar2016a}%
  \BibitemOpen
  \bibfield  {author} {\bibinfo {author} {\bibfnamefont {P.}~\bibnamefont
  {Kumar}}, \bibinfo {author} {\bibfnamefont {S.}~\bibnamefont {Sendelbach}},
  \bibinfo {author} {\bibfnamefont {M.}~\bibnamefont {Beck}}, \bibinfo {author}
  {\bibfnamefont {J.}~\bibnamefont {Freeland}}, \bibinfo {author}
  {\bibfnamefont {Z.}~\bibnamefont {Wang}}, \bibinfo {author} {\bibfnamefont
  {H.}~\bibnamefont {Wang}}, \bibinfo {author} {\bibfnamefont {C.~C.}\
  \bibnamefont {Yu}}, \bibinfo {author} {\bibfnamefont {R.}~\bibnamefont {Wu}},
  \bibinfo {author} {\bibfnamefont {D.}~\bibnamefont {Pappas}}, \ and\ \bibinfo
  {author} {\bibfnamefont {R.}~\bibnamefont {McDermott}},\ }\href {\doibase
  10.1103/PhysRevApplied.6.041001} {\bibfield  {journal} {\bibinfo  {journal}
  {Physical Review Applied}\ }\textbf {\bibinfo {volume} {6}},\ \bibinfo
  {pages} {041001} (\bibinfo {year} {2016})}\BibitemShut {NoStop}%
\bibitem [{\citenamefont {McGuirk}\ \emph {et~al.}(2004)\citenamefont
  {McGuirk}, \citenamefont {Harber}, \citenamefont {Obrecht},\ and\
  \citenamefont {Cornell}}]{McGuirk2004}%
  \BibitemOpen
  \bibfield  {author} {\bibinfo {author} {\bibfnamefont {J.~M.}\ \bibnamefont
  {McGuirk}}, \bibinfo {author} {\bibfnamefont {D.~M.}\ \bibnamefont {Harber}},
  \bibinfo {author} {\bibfnamefont {J.~M.}\ \bibnamefont {Obrecht}}, \ and\
  \bibinfo {author} {\bibfnamefont {E.~A.}\ \bibnamefont {Cornell}},\ }\href
  {\doibase 10.1103/PhysRevA.69.062905} {\bibfield  {journal} {\bibinfo
  {journal} {Physical Review A}\ }\textbf {\bibinfo {volume} {69}},\ \bibinfo
  {pages} {062905} (\bibinfo {year} {2004})}\BibitemShut {NoStop}%
\bibitem [{\citenamefont {Daniilidis}\ \emph {et~al.}(2011)\citenamefont
  {Daniilidis}, \citenamefont {Narayanan}, \citenamefont {M{\"{o}}ller},
  \citenamefont {Clark}, \citenamefont {Lee}, \citenamefont {Leek},
  \citenamefont {Wallraff}, \citenamefont {Schulz}, \citenamefont
  {Schmidt-Kaler},\ and\ \citenamefont {H{\"{a}}ffner}}]{Daniilidis2011}%
  \BibitemOpen
  \bibfield  {author} {\bibinfo {author} {\bibfnamefont {N.}~\bibnamefont
  {Daniilidis}}, \bibinfo {author} {\bibfnamefont {S.}~\bibnamefont
  {Narayanan}}, \bibinfo {author} {\bibfnamefont {S.~A.}\ \bibnamefont
  {M{\"{o}}ller}}, \bibinfo {author} {\bibfnamefont {R.}~\bibnamefont {Clark}},
  \bibinfo {author} {\bibfnamefont {T.~E.}\ \bibnamefont {Lee}}, \bibinfo
  {author} {\bibfnamefont {P.~J.}\ \bibnamefont {Leek}}, \bibinfo {author}
  {\bibfnamefont {A.}~\bibnamefont {Wallraff}}, \bibinfo {author}
  {\bibfnamefont {S.}~\bibnamefont {Schulz}}, \bibinfo {author} {\bibfnamefont
  {F.}~\bibnamefont {Schmidt-Kaler}}, \ and\ \bibinfo {author} {\bibfnamefont
  {H.}~\bibnamefont {H{\"{a}}ffner}},\ }\href {\doibase
  10.1088/1367-2630/13/1/013032} {\bibfield  {journal} {\bibinfo  {journal}
  {New Journal of Physics}\ }\textbf {\bibinfo {volume} {13}},\ \bibinfo
  {pages} {013032} (\bibinfo {year} {2011})}\BibitemShut {NoStop}%
\bibitem [{\citenamefont {Hite}\ \emph {et~al.}(2012)\citenamefont {Hite},
  \citenamefont {Colombe}, \citenamefont {Wilson}, \citenamefont {Brown},
  \citenamefont {Warring}, \citenamefont {J{\"{o}}rdens}, \citenamefont {Jost},
  \citenamefont {McKay}, \citenamefont {Pappas}, \citenamefont {Leibfried},\
  and\ \citenamefont {Wineland}}]{Hite2012}%
  \BibitemOpen
  \bibfield  {author} {\bibinfo {author} {\bibfnamefont {D.~A.}\ \bibnamefont
  {Hite}}, \bibinfo {author} {\bibfnamefont {Y.}~\bibnamefont {Colombe}},
  \bibinfo {author} {\bibfnamefont {A.~C.}\ \bibnamefont {Wilson}}, \bibinfo
  {author} {\bibfnamefont {K.~R.}\ \bibnamefont {Brown}}, \bibinfo {author}
  {\bibfnamefont {U.}~\bibnamefont {Warring}}, \bibinfo {author} {\bibfnamefont
  {R.}~\bibnamefont {J{\"{o}}rdens}}, \bibinfo {author} {\bibfnamefont {J.~D.}\
  \bibnamefont {Jost}}, \bibinfo {author} {\bibfnamefont {K.~S.}\ \bibnamefont
  {McKay}}, \bibinfo {author} {\bibfnamefont {D.~P.}\ \bibnamefont {Pappas}},
  \bibinfo {author} {\bibfnamefont {D.}~\bibnamefont {Leibfried}}, \ and\
  \bibinfo {author} {\bibfnamefont {D.~J.}\ \bibnamefont {Wineland}},\ }\href
  {\doibase 10.1103/PhysRevLett.109.103001} {\bibfield  {journal} {\bibinfo
  {journal} {Physical Review Letters}\ }\textbf {\bibinfo {volume} {109}},\
  \bibinfo {pages} {103001} (\bibinfo {year} {2012})}\BibitemShut {NoStop}%
\bibitem [{\citenamefont {Bluvstein}\ \emph {et~al.}(2019)\citenamefont
  {Bluvstein}, \citenamefont {Zhang},\ and\ \citenamefont
  {Jayich}}]{Bluvstein2019}%
  \BibitemOpen
  \bibfield  {author} {\bibinfo {author} {\bibfnamefont {D.}~\bibnamefont
  {Bluvstein}}, \bibinfo {author} {\bibfnamefont {Z.}~\bibnamefont {Zhang}}, \
  and\ \bibinfo {author} {\bibfnamefont {A.~C.~B.}\ \bibnamefont {Jayich}},\
  }\href {\doibase 10.1103/PhysRevLett.122.076101} {\bibfield  {journal}
  {\bibinfo  {journal} {Physical Review Letters}\ }\textbf {\bibinfo {volume}
  {122}},\ \bibinfo {pages} {076101} (\bibinfo {year} {2019})}\BibitemShut
  {NoStop}%
\bibitem [{\citenamefont {Du}\ \emph {et~al.}(2009)\citenamefont {Du},
  \citenamefont {Rong}, \citenamefont {Zhao}, \citenamefont {Wang},
  \citenamefont {Yang},\ and\ \citenamefont {Liu}}]{Du2009}%
  \BibitemOpen
  \bibfield  {author} {\bibinfo {author} {\bibfnamefont {J.}~\bibnamefont
  {Du}}, \bibinfo {author} {\bibfnamefont {X.}~\bibnamefont {Rong}}, \bibinfo
  {author} {\bibfnamefont {N.}~\bibnamefont {Zhao}}, \bibinfo {author}
  {\bibfnamefont {Y.}~\bibnamefont {Wang}}, \bibinfo {author} {\bibfnamefont
  {J.}~\bibnamefont {Yang}}, \ and\ \bibinfo {author} {\bibfnamefont {R.~B.}\
  \bibnamefont {Liu}},\ }\href {\doibase 10.1038/nature08470} {\bibfield
  {journal} {\bibinfo  {journal} {Nature}\ }\textbf {\bibinfo {volume} {461}},\
  \bibinfo {pages} {1265} (\bibinfo {year} {2009})}\BibitemShut {NoStop}%
\bibitem [{\citenamefont {de~Lange}\ \emph {et~al.}(2010)\citenamefont
  {de~Lange}, \citenamefont {Wang}, \citenamefont {Rist{\`{e}}}, \citenamefont
  {Dobrovitski},\ and\ \citenamefont {Hanson}}]{DeLange2010}%
  \BibitemOpen
  \bibfield  {author} {\bibinfo {author} {\bibfnamefont {G.}~\bibnamefont
  {de~Lange}}, \bibinfo {author} {\bibfnamefont {Z.~H.}\ \bibnamefont {Wang}},
  \bibinfo {author} {\bibfnamefont {D.}~\bibnamefont {Rist{\`{e}}}}, \bibinfo
  {author} {\bibfnamefont {V.~V.}\ \bibnamefont {Dobrovitski}}, \ and\ \bibinfo
  {author} {\bibfnamefont {R.}~\bibnamefont {Hanson}},\ }\href {\doibase
  10.1126/science.1192739} {\bibfield  {journal} {\bibinfo  {journal} {Science
  (New York, N.Y.)}\ }\textbf {\bibinfo {volume} {330}},\ \bibinfo {pages} {60}
  (\bibinfo {year} {2010})}\BibitemShut {NoStop}%
\bibitem [{\citenamefont {Bylander}\ \emph {et~al.}(2011)\citenamefont
  {Bylander}, \citenamefont {Gustavsson}, \citenamefont {Yan}, \citenamefont
  {Yoshihara}, \citenamefont {Harrabi}, \citenamefont {Fitch}, \citenamefont
  {Cory}, \citenamefont {Nakamura}, \citenamefont {Tsai},\ and\ \citenamefont
  {Oliver}}]{Bylander2011}%
  \BibitemOpen
  \bibfield  {author} {\bibinfo {author} {\bibfnamefont {J.}~\bibnamefont
  {Bylander}}, \bibinfo {author} {\bibfnamefont {S.}~\bibnamefont
  {Gustavsson}}, \bibinfo {author} {\bibfnamefont {F.}~\bibnamefont {Yan}},
  \bibinfo {author} {\bibfnamefont {F.}~\bibnamefont {Yoshihara}}, \bibinfo
  {author} {\bibfnamefont {K.}~\bibnamefont {Harrabi}}, \bibinfo {author}
  {\bibfnamefont {G.}~\bibnamefont {Fitch}}, \bibinfo {author} {\bibfnamefont
  {D.~G.}\ \bibnamefont {Cory}}, \bibinfo {author} {\bibfnamefont
  {Y.}~\bibnamefont {Nakamura}}, \bibinfo {author} {\bibfnamefont {J.-S.}\
  \bibnamefont {Tsai}}, \ and\ \bibinfo {author} {\bibfnamefont {W.~D.}\
  \bibnamefont {Oliver}},\ }\href {\doibase 10.1038/nphys1994} {\bibfield
  {journal} {\bibinfo  {journal} {Nature Physics}\ }\textbf {\bibinfo {volume}
  {7}},\ \bibinfo {pages} {565} (\bibinfo {year} {2011})}\BibitemShut {NoStop}%
\bibitem [{\citenamefont {Koch}\ \emph {et~al.}(2007)\citenamefont {Koch},
  \citenamefont {Yu}, \citenamefont {Gambetta}, \citenamefont {Houck},
  \citenamefont {Schuster}, \citenamefont {Majer}, \citenamefont {Blais},
  \citenamefont {Devoret}, \citenamefont {Girvin},\ and\ \citenamefont
  {Schoelkopf}}]{Koch2007}%
  \BibitemOpen
  \bibfield  {author} {\bibinfo {author} {\bibfnamefont {J.}~\bibnamefont
  {Koch}}, \bibinfo {author} {\bibfnamefont {T.~M.}\ \bibnamefont {Yu}},
  \bibinfo {author} {\bibfnamefont {J.}~\bibnamefont {Gambetta}}, \bibinfo
  {author} {\bibfnamefont {A.~A.}\ \bibnamefont {Houck}}, \bibinfo {author}
  {\bibfnamefont {D.~I.}\ \bibnamefont {Schuster}}, \bibinfo {author}
  {\bibfnamefont {J.}~\bibnamefont {Majer}}, \bibinfo {author} {\bibfnamefont
  {A.}~\bibnamefont {Blais}}, \bibinfo {author} {\bibfnamefont {M.~H.}\
  \bibnamefont {Devoret}}, \bibinfo {author} {\bibfnamefont {S.~M.}\
  \bibnamefont {Girvin}}, \ and\ \bibinfo {author} {\bibfnamefont {R.~J.}\
  \bibnamefont {Schoelkopf}},\ }\href {\doibase 10.1103/PhysRevA.76.042319}
  {\bibfield  {journal} {\bibinfo  {journal} {Physical Review A}\ }\textbf
  {\bibinfo {volume} {76}},\ \bibinfo {pages} {042319} (\bibinfo {year}
  {2007})}\BibitemShut {NoStop}%
\bibitem [{\citenamefont {Wolfowicz}\ \emph {et~al.}(2013)\citenamefont
  {Wolfowicz}, \citenamefont {Tyryshkin}, \citenamefont {George}, \citenamefont
  {Riemann}, \citenamefont {Abrosimov}, \citenamefont {Becker}, \citenamefont
  {Pohl}, \citenamefont {Thewalt}, \citenamefont {Lyon},\ and\ \citenamefont
  {Morton}}]{Wolfowicz2013}%
  \BibitemOpen
  \bibfield  {author} {\bibinfo {author} {\bibfnamefont {G.}~\bibnamefont
  {Wolfowicz}}, \bibinfo {author} {\bibfnamefont {A.~M.}\ \bibnamefont
  {Tyryshkin}}, \bibinfo {author} {\bibfnamefont {R.~E.}\ \bibnamefont
  {George}}, \bibinfo {author} {\bibfnamefont {H.}~\bibnamefont {Riemann}},
  \bibinfo {author} {\bibfnamefont {N.~V.}\ \bibnamefont {Abrosimov}}, \bibinfo
  {author} {\bibfnamefont {P.}~\bibnamefont {Becker}}, \bibinfo {author}
  {\bibfnamefont {H.-J.}\ \bibnamefont {Pohl}}, \bibinfo {author}
  {\bibfnamefont {M.~L.~W.}\ \bibnamefont {Thewalt}}, \bibinfo {author}
  {\bibfnamefont {S.~A.}\ \bibnamefont {Lyon}}, \ and\ \bibinfo {author}
  {\bibfnamefont {J.~J.~L.}\ \bibnamefont {Morton}},\ }\href {\doibase
  10.1038/nnano.2013.117} {\bibfield  {journal} {\bibinfo  {journal} {Nature
  Nanotechnology}\ }\textbf {\bibinfo {volume} {8}},\ \bibinfo {pages} {561}
  (\bibinfo {year} {2013})}\BibitemShut {NoStop}%
\bibitem [{\citenamefont {Zadrozny}\ \emph {et~al.}(2017)\citenamefont
  {Zadrozny}, \citenamefont {Gallagher}, \citenamefont {Harris},\ and\
  \citenamefont {Freedman}}]{Zadrozny2017}%
  \BibitemOpen
  \bibfield  {author} {\bibinfo {author} {\bibfnamefont {J.~M.}\ \bibnamefont
  {Zadrozny}}, \bibinfo {author} {\bibfnamefont {A.~T.}\ \bibnamefont
  {Gallagher}}, \bibinfo {author} {\bibfnamefont {T.~D.}\ \bibnamefont
  {Harris}}, \ and\ \bibinfo {author} {\bibfnamefont {D.~E.}\ \bibnamefont
  {Freedman}},\ }\href {\doibase 10.1021/jacs.7b03123} {\bibfield  {journal}
  {\bibinfo  {journal} {Journal of the American Chemical Society}\ }\textbf
  {\bibinfo {volume} {139}},\ \bibinfo {pages} {7089} (\bibinfo {year}
  {2017})}\BibitemShut {NoStop}%
\bibitem [{\citenamefont {Barfuss}\ \emph {et~al.}(2018)\citenamefont
  {Barfuss}, \citenamefont {K{\"{o}}lbl}, \citenamefont {Thiel}, \citenamefont
  {Teissier}, \citenamefont {Kasperczyk},\ and\ \citenamefont
  {Maletinsky}}]{Barfuss2018}%
  \BibitemOpen
  \bibfield  {author} {\bibinfo {author} {\bibfnamefont {A.}~\bibnamefont
  {Barfuss}}, \bibinfo {author} {\bibfnamefont {J.}~\bibnamefont
  {K{\"{o}}lbl}}, \bibinfo {author} {\bibfnamefont {L.}~\bibnamefont {Thiel}},
  \bibinfo {author} {\bibfnamefont {J.}~\bibnamefont {Teissier}}, \bibinfo
  {author} {\bibfnamefont {M.}~\bibnamefont {Kasperczyk}}, \ and\ \bibinfo
  {author} {\bibfnamefont {P.}~\bibnamefont {Maletinsky}},\ }\href {\doibase
  10.1038/s41567-018-0231-8} {\bibfield  {journal} {\bibinfo  {journal} {Nature
  Physics}\ }\textbf {\bibinfo {volume} {14}},\ \bibinfo {pages} {1087}
  (\bibinfo {year} {2018})}\BibitemShut {NoStop}%
\bibitem [{\citenamefont {Slichter}(1990)}]{Slichter1990}%
  \BibitemOpen
  \bibfield  {author} {\bibinfo {author} {\bibfnamefont {C.~P.}\ \bibnamefont
  {Slichter}},\ }\href@noop {} {\emph {\bibinfo {title} {{Principles of
  magnetic resonance}}}}\ (\bibinfo  {publisher} {Springer-Verlag},\ \bibinfo
  {year} {1990})\BibitemShut {NoStop}%
\bibitem [{\citenamefont {Sangtawesin}\ \emph {et~al.}(2018)\citenamefont
  {Sangtawesin}, \citenamefont {Dwyer}, \citenamefont {Srinivasan},
  \citenamefont {Allred}, \citenamefont {Rodgers}, \citenamefont {{De Greve}},
  \citenamefont {Stacey}, \citenamefont {Dontschuk}, \citenamefont {O'Donnell},
  \citenamefont {Hu}, \citenamefont {Evans}, \citenamefont {Jaye},
  \citenamefont {Fischer}, \citenamefont {Markham}, \citenamefont {Twitchen},
  \citenamefont {Park}, \citenamefont {Lukin},\ and\ \citenamefont
  {de~Leon}}]{Sangtawesin2018}%
  \BibitemOpen
  \bibfield  {author} {\bibinfo {author} {\bibfnamefont {S.}~\bibnamefont
  {Sangtawesin}}, \bibinfo {author} {\bibfnamefont {B.~L.}\ \bibnamefont
  {Dwyer}}, \bibinfo {author} {\bibfnamefont {S.}~\bibnamefont {Srinivasan}},
  \bibinfo {author} {\bibfnamefont {J.~J.}\ \bibnamefont {Allred}}, \bibinfo
  {author} {\bibfnamefont {L.~V.~H.}\ \bibnamefont {Rodgers}}, \bibinfo
  {author} {\bibfnamefont {K.}~\bibnamefont {{De Greve}}}, \bibinfo {author}
  {\bibfnamefont {A.}~\bibnamefont {Stacey}}, \bibinfo {author} {\bibfnamefont
  {N.}~\bibnamefont {Dontschuk}}, \bibinfo {author} {\bibfnamefont {K.~M.}\
  \bibnamefont {O'Donnell}}, \bibinfo {author} {\bibfnamefont {D.}~\bibnamefont
  {Hu}}, \bibinfo {author} {\bibfnamefont {D.~A.}\ \bibnamefont {Evans}},
  \bibinfo {author} {\bibfnamefont {C.}~\bibnamefont {Jaye}}, \bibinfo {author}
  {\bibfnamefont {D.~A.}\ \bibnamefont {Fischer}}, \bibinfo {author}
  {\bibfnamefont {M.~L.}\ \bibnamefont {Markham}}, \bibinfo {author}
  {\bibfnamefont {D.~J.}\ \bibnamefont {Twitchen}}, \bibinfo {author}
  {\bibfnamefont {H.}~\bibnamefont {Park}}, \bibinfo {author} {\bibfnamefont
  {M.~D.}\ \bibnamefont {Lukin}}, \ and\ \bibinfo {author} {\bibfnamefont
  {N.~P.}\ \bibnamefont {de~Leon}},\ }\href {http://arxiv.org/abs/1811.00144}
  {\  (\bibinfo {year} {2018})},\ \Eprint {http://arxiv.org/abs/1811.00144}
  {arXiv:1811.00144} \BibitemShut {NoStop}%
\bibitem [{Note1()}]{Note1}%
  \BibitemOpen
  \bibinfo {note} {See the Supplemental Material for supporting details on
  sample information, surface spin density, RF control, the AC Stark effect,
  and DQ and SQ susceptibilities, which includes Refs.~\cite
  {Toyli2010,DeWit2018,Friedrich2017}}\BibitemShut {NoStop}%
\bibitem [{\citenamefont {Pham}\ \emph {et~al.}(2016)\citenamefont {Pham},
  \citenamefont {DeVience}, \citenamefont {Casola}, \citenamefont {Lovchinsky},
  \citenamefont {Sushkov}, \citenamefont {Bersin}, \citenamefont {Lee},
  \citenamefont {Urbach}, \citenamefont {Cappellaro}, \citenamefont {Park},
  \citenamefont {Yacoby}, \citenamefont {Lukin},\ and\ \citenamefont
  {Walsworth}}]{Pham2016}%
  \BibitemOpen
  \bibfield  {author} {\bibinfo {author} {\bibfnamefont {L.~M.}\ \bibnamefont
  {Pham}}, \bibinfo {author} {\bibfnamefont {S.~J.}\ \bibnamefont {DeVience}},
  \bibinfo {author} {\bibfnamefont {F.}~\bibnamefont {Casola}}, \bibinfo
  {author} {\bibfnamefont {I.}~\bibnamefont {Lovchinsky}}, \bibinfo {author}
  {\bibfnamefont {A.~O.}\ \bibnamefont {Sushkov}}, \bibinfo {author}
  {\bibfnamefont {E.}~\bibnamefont {Bersin}}, \bibinfo {author} {\bibfnamefont
  {J.}~\bibnamefont {Lee}}, \bibinfo {author} {\bibfnamefont {E.}~\bibnamefont
  {Urbach}}, \bibinfo {author} {\bibfnamefont {P.}~\bibnamefont {Cappellaro}},
  \bibinfo {author} {\bibfnamefont {H.}~\bibnamefont {Park}}, \bibinfo {author}
  {\bibfnamefont {A.}~\bibnamefont {Yacoby}}, \bibinfo {author} {\bibfnamefont
  {M.}~\bibnamefont {Lukin}}, \ and\ \bibinfo {author} {\bibfnamefont {R.~L.}\
  \bibnamefont {Walsworth}},\ }\href {\doibase 10.1103/PhysRevB.93.045425}
  {\bibfield  {journal} {\bibinfo  {journal} {Physical Review B}\ }\textbf
  {\bibinfo {volume} {93}},\ \bibinfo {pages} {045425} (\bibinfo {year}
  {2016})}\BibitemShut {NoStop}%
\bibitem [{\citenamefont {Mamin}\ \emph {et~al.}(2012)\citenamefont {Mamin},
  \citenamefont {Sherwood},\ and\ \citenamefont {Rugar}}]{Mamin2012}%
  \BibitemOpen
  \bibfield  {author} {\bibinfo {author} {\bibfnamefont {H.~J.}\ \bibnamefont
  {Mamin}}, \bibinfo {author} {\bibfnamefont {M.~H.}\ \bibnamefont {Sherwood}},
  \ and\ \bibinfo {author} {\bibfnamefont {D.}~\bibnamefont {Rugar}},\ }\href
  {\doibase 10.1103/PhysRevB.86.195422} {\bibfield  {journal} {\bibinfo
  {journal} {Physical Review B}\ }\textbf {\bibinfo {volume} {86}},\ \bibinfo
  {pages} {195422} (\bibinfo {year} {2012})}\BibitemShut {NoStop}%
\bibitem [{\citenamefont {Grinolds}\ \emph {et~al.}(2014)\citenamefont
  {Grinolds}, \citenamefont {Warner}, \citenamefont {{De Greve}}, \citenamefont
  {Dovzhenko}, \citenamefont {Thiel}, \citenamefont {Walsworth}, \citenamefont
  {Hong}, \citenamefont {Maletinsky},\ and\ \citenamefont
  {Yacoby}}]{Grinolds2014}%
  \BibitemOpen
  \bibfield  {author} {\bibinfo {author} {\bibfnamefont {M.~S.}\ \bibnamefont
  {Grinolds}}, \bibinfo {author} {\bibfnamefont {M.}~\bibnamefont {Warner}},
  \bibinfo {author} {\bibfnamefont {K.}~\bibnamefont {{De Greve}}}, \bibinfo
  {author} {\bibfnamefont {Y.}~\bibnamefont {Dovzhenko}}, \bibinfo {author}
  {\bibfnamefont {L.}~\bibnamefont {Thiel}}, \bibinfo {author} {\bibfnamefont
  {R.~L.}\ \bibnamefont {Walsworth}}, \bibinfo {author} {\bibfnamefont
  {S.}~\bibnamefont {Hong}}, \bibinfo {author} {\bibfnamefont {P.}~\bibnamefont
  {Maletinsky}}, \ and\ \bibinfo {author} {\bibfnamefont {A.}~\bibnamefont
  {Yacoby}},\ }\href {\doibase 10.1038/nnano.2014.30} {\bibfield  {journal}
  {\bibinfo  {journal} {Nature Nanotechnology}\ }\textbf {\bibinfo {volume}
  {9}},\ \bibinfo {pages} {279} (\bibinfo {year} {2014})}\BibitemShut {NoStop}%
\bibitem [{\citenamefont {Sushkov}\ \emph {et~al.}(2014)\citenamefont
  {Sushkov}, \citenamefont {Lovchinsky}, \citenamefont {Chisholm},
  \citenamefont {Walsworth}, \citenamefont {Park},\ and\ \citenamefont
  {Lukin}}]{Sushkov2014}%
  \BibitemOpen
  \bibfield  {author} {\bibinfo {author} {\bibfnamefont {A.}~\bibnamefont
  {Sushkov}}, \bibinfo {author} {\bibfnamefont {I.}~\bibnamefont {Lovchinsky}},
  \bibinfo {author} {\bibfnamefont {N.}~\bibnamefont {Chisholm}}, \bibinfo
  {author} {\bibfnamefont {R.}~\bibnamefont {Walsworth}}, \bibinfo {author}
  {\bibfnamefont {H.}~\bibnamefont {Park}}, \ and\ \bibinfo {author}
  {\bibfnamefont {M.}~\bibnamefont {Lukin}},\ }\href {\doibase
  10.1103/PhysRevLett.113.197601} {\bibfield  {journal} {\bibinfo  {journal}
  {Physical Review Letters}\ }\textbf {\bibinfo {volume} {113}},\ \bibinfo
  {pages} {197601} (\bibinfo {year} {2014})}\BibitemShut {NoStop}%
\bibitem [{\citenamefont {Mamin}\ \emph {et~al.}(2014)\citenamefont {Mamin},
  \citenamefont {Sherwood}, \citenamefont {Kim}, \citenamefont {Rettner},
  \citenamefont {Ohno}, \citenamefont {Awschalom},\ and\ \citenamefont
  {Rugar}}]{Mamin2014}%
  \BibitemOpen
  \bibfield  {author} {\bibinfo {author} {\bibfnamefont {H.}~\bibnamefont
  {Mamin}}, \bibinfo {author} {\bibfnamefont {M.}~\bibnamefont {Sherwood}},
  \bibinfo {author} {\bibfnamefont {M.}~\bibnamefont {Kim}}, \bibinfo {author}
  {\bibfnamefont {C.}~\bibnamefont {Rettner}}, \bibinfo {author} {\bibfnamefont
  {K.}~\bibnamefont {Ohno}}, \bibinfo {author} {\bibfnamefont {D.}~\bibnamefont
  {Awschalom}}, \ and\ \bibinfo {author} {\bibfnamefont {D.}~\bibnamefont
  {Rugar}},\ }\href {\doibase 10.1103/PhysRevLett.113.030803} {\bibfield
  {journal} {\bibinfo  {journal} {Physical Review Letters}\ }\textbf {\bibinfo
  {volume} {113}},\ \bibinfo {pages} {030803} (\bibinfo {year}
  {2014})}\BibitemShut {NoStop}%
\bibitem [{\citenamefont {Bauch}\ \emph {et~al.}(2018)\citenamefont {Bauch},
  \citenamefont {Hart}, \citenamefont {Schloss}, \citenamefont {Turner},
  \citenamefont {Barry}, \citenamefont {Kehayias}, \citenamefont {Singh},\ and\
  \citenamefont {Walsworth}}]{Bauch2018}%
  \BibitemOpen
  \bibfield  {author} {\bibinfo {author} {\bibfnamefont {E.}~\bibnamefont
  {Bauch}}, \bibinfo {author} {\bibfnamefont {C.~A.}\ \bibnamefont {Hart}},
  \bibinfo {author} {\bibfnamefont {J.~M.}\ \bibnamefont {Schloss}}, \bibinfo
  {author} {\bibfnamefont {M.~J.}\ \bibnamefont {Turner}}, \bibinfo {author}
  {\bibfnamefont {J.~F.}\ \bibnamefont {Barry}}, \bibinfo {author}
  {\bibfnamefont {P.}~\bibnamefont {Kehayias}}, \bibinfo {author}
  {\bibfnamefont {S.}~\bibnamefont {Singh}}, \ and\ \bibinfo {author}
  {\bibfnamefont {R.~L.}\ \bibnamefont {Walsworth}},\ }\href {\doibase
  10.1103/PhysRevX.8.031025} {\bibfield  {journal} {\bibinfo  {journal}
  {Physical Review X}\ }\textbf {\bibinfo {volume} {8}},\ \bibinfo {pages}
  {031025} (\bibinfo {year} {2018})}\BibitemShut {NoStop}%
\bibitem [{\citenamefont {Hall}\ \emph {et~al.}(2009)\citenamefont {Hall},
  \citenamefont {Cole}, \citenamefont {Hill},\ and\ \citenamefont
  {Hollenberg}}]{Hall2009}%
  \BibitemOpen
  \bibfield  {author} {\bibinfo {author} {\bibfnamefont {L.~T.}\ \bibnamefont
  {Hall}}, \bibinfo {author} {\bibfnamefont {J.~H.}\ \bibnamefont {Cole}},
  \bibinfo {author} {\bibfnamefont {C.~D.}\ \bibnamefont {Hill}}, \ and\
  \bibinfo {author} {\bibfnamefont {L.~C.~L.}\ \bibnamefont {Hollenberg}},\
  }\href {\doibase 10.1103/PhysRevLett.103.220802} {\bibfield  {journal}
  {\bibinfo  {journal} {Physical Review Letters}\ }\textbf {\bibinfo {volume}
  {103}},\ \bibinfo {pages} {220802} (\bibinfo {year} {2009})}\BibitemShut
  {NoStop}%
\bibitem [{\citenamefont {Sch{\"{a}}fer-Nolte}\ \emph
  {et~al.}(2014)\citenamefont {Sch{\"{a}}fer-Nolte}, \citenamefont {Schlipf},
  \citenamefont {Ternes}, \citenamefont {Reinhard}, \citenamefont {Kern},\ and\
  \citenamefont {Wrachtrup}}]{Schafer-Nolte2014a}%
  \BibitemOpen
  \bibfield  {author} {\bibinfo {author} {\bibfnamefont {E.}~\bibnamefont
  {Sch{\"{a}}fer-Nolte}}, \bibinfo {author} {\bibfnamefont {L.}~\bibnamefont
  {Schlipf}}, \bibinfo {author} {\bibfnamefont {M.}~\bibnamefont {Ternes}},
  \bibinfo {author} {\bibfnamefont {F.}~\bibnamefont {Reinhard}}, \bibinfo
  {author} {\bibfnamefont {K.}~\bibnamefont {Kern}}, \ and\ \bibinfo {author}
  {\bibfnamefont {J.}~\bibnamefont {Wrachtrup}},\ }\href {\doibase
  10.1103/PhysRevLett.113.217204} {\bibfield  {journal} {\bibinfo  {journal}
  {Physical Review Letters}\ }\textbf {\bibinfo {volume} {113}},\ \bibinfo
  {pages} {217204} (\bibinfo {year} {2014})}\BibitemShut {NoStop}%
\bibitem [{\citenamefont {Meriles}\ \emph {et~al.}(2010)\citenamefont
  {Meriles}, \citenamefont {Jiang}, \citenamefont {Goldstein}, \citenamefont
  {Hodges}, \citenamefont {Maze}, \citenamefont {Lukin},\ and\ \citenamefont
  {Cappellaro}}]{Meriles2010}%
  \BibitemOpen
  \bibfield  {author} {\bibinfo {author} {\bibfnamefont {C.~A.}\ \bibnamefont
  {Meriles}}, \bibinfo {author} {\bibfnamefont {L.}~\bibnamefont {Jiang}},
  \bibinfo {author} {\bibfnamefont {G.}~\bibnamefont {Goldstein}}, \bibinfo
  {author} {\bibfnamefont {J.~S.}\ \bibnamefont {Hodges}}, \bibinfo {author}
  {\bibfnamefont {J.}~\bibnamefont {Maze}}, \bibinfo {author} {\bibfnamefont
  {M.~D.}\ \bibnamefont {Lukin}}, \ and\ \bibinfo {author} {\bibfnamefont
  {P.}~\bibnamefont {Cappellaro}},\ }\href {\doibase 10.1063/1.3483676}
  {\bibfield  {journal} {\bibinfo  {journal} {The Journal of Chemical Physics}\
  }\textbf {\bibinfo {volume} {133}},\ \bibinfo {pages} {124105} (\bibinfo
  {year} {2010})}\BibitemShut {NoStop}%
\bibitem [{\citenamefont {{F{\'{a}}varo de Oliveira}}\ \emph
  {et~al.}(2017)\citenamefont {{F{\'{a}}varo de Oliveira}}, \citenamefont
  {Antonov}, \citenamefont {Wang}, \citenamefont {Neumann}, \citenamefont
  {Momenzadeh}, \citenamefont {H{\"{a}}u{\ss}ermann}, \citenamefont
  {Pasquarelli}, \citenamefont {Denisenko},\ and\ \citenamefont
  {Wrachtrup}}]{FavarodeOliveira2017a}%
  \BibitemOpen
  \bibfield  {author} {\bibinfo {author} {\bibfnamefont {F.}~\bibnamefont
  {{F{\'{a}}varo de Oliveira}}}, \bibinfo {author} {\bibfnamefont
  {D.}~\bibnamefont {Antonov}}, \bibinfo {author} {\bibfnamefont
  {Y.}~\bibnamefont {Wang}}, \bibinfo {author} {\bibfnamefont {P.}~\bibnamefont
  {Neumann}}, \bibinfo {author} {\bibfnamefont {S.~A.}\ \bibnamefont
  {Momenzadeh}}, \bibinfo {author} {\bibfnamefont {T.}~\bibnamefont
  {H{\"{a}}u{\ss}ermann}}, \bibinfo {author} {\bibfnamefont {A.}~\bibnamefont
  {Pasquarelli}}, \bibinfo {author} {\bibfnamefont {A.}~\bibnamefont
  {Denisenko}}, \ and\ \bibinfo {author} {\bibfnamefont {J.}~\bibnamefont
  {Wrachtrup}},\ }\href {\doibase 10.1038/ncomms15409} {\bibfield  {journal}
  {\bibinfo  {journal} {Nature Communications}\ }\textbf {\bibinfo {volume}
  {8}},\ \bibinfo {pages} {15409} (\bibinfo {year} {2017})}\BibitemShut
  {NoStop}%
\bibitem [{\citenamefont {Bauch}\ \emph {et~al.}(2019)\citenamefont {Bauch},
  \citenamefont {Singh}, \citenamefont {Lee}, \citenamefont {Hart},
  \citenamefont {Schloss}, \citenamefont {Turner}, \citenamefont {Barry},
  \citenamefont {Pham}, \citenamefont {Bar-Gill}, \citenamefont {Yelin},\ and\
  \citenamefont {Walsworth}}]{Bauch2019}%
  \BibitemOpen
  \bibfield  {author} {\bibinfo {author} {\bibfnamefont {E.}~\bibnamefont
  {Bauch}}, \bibinfo {author} {\bibfnamefont {S.}~\bibnamefont {Singh}},
  \bibinfo {author} {\bibfnamefont {J.}~\bibnamefont {Lee}}, \bibinfo {author}
  {\bibfnamefont {C.~A.}\ \bibnamefont {Hart}}, \bibinfo {author}
  {\bibfnamefont {J.~M.}\ \bibnamefont {Schloss}}, \bibinfo {author}
  {\bibfnamefont {M.~J.}\ \bibnamefont {Turner}}, \bibinfo {author}
  {\bibfnamefont {J.~F.}\ \bibnamefont {Barry}}, \bibinfo {author}
  {\bibfnamefont {L.}~\bibnamefont {Pham}}, \bibinfo {author} {\bibfnamefont
  {N.}~\bibnamefont {Bar-Gill}}, \bibinfo {author} {\bibfnamefont {S.~F.}\
  \bibnamefont {Yelin}}, \ and\ \bibinfo {author} {\bibfnamefont {R.~L.}\
  \bibnamefont {Walsworth}},\ }\href {http://arxiv.org/abs/1904.08763} {\
  (\bibinfo {year} {2019})},\ \Eprint {http://arxiv.org/abs/1904.08763}
  {arXiv:1904.08763} \BibitemShut {NoStop}%
\bibitem [{\citenamefont {de~Lange}\ \emph {et~al.}(2012)\citenamefont
  {de~Lange}, \citenamefont {van~der Sar}, \citenamefont {Blok}, \citenamefont
  {Wang}, \citenamefont {Dobrovitski},\ and\ \citenamefont
  {Hanson}}]{DeLange2012}%
  \BibitemOpen
  \bibfield  {author} {\bibinfo {author} {\bibfnamefont {G.}~\bibnamefont
  {de~Lange}}, \bibinfo {author} {\bibfnamefont {T.}~\bibnamefont {van~der
  Sar}}, \bibinfo {author} {\bibfnamefont {M.}~\bibnamefont {Blok}}, \bibinfo
  {author} {\bibfnamefont {Z.-H.}\ \bibnamefont {Wang}}, \bibinfo {author}
  {\bibfnamefont {V.}~\bibnamefont {Dobrovitski}}, \ and\ \bibinfo {author}
  {\bibfnamefont {R.}~\bibnamefont {Hanson}},\ }\href {\doibase
  10.1038/srep00382} {\bibfield  {journal} {\bibinfo  {journal} {Scientific
  Reports}\ }\textbf {\bibinfo {volume} {2}},\ \bibinfo {pages} {382} (\bibinfo
  {year} {2012})}\BibitemShut {NoStop}%
\bibitem [{\citenamefont {Knowles}\ \emph {et~al.}(2014)\citenamefont
  {Knowles}, \citenamefont {Kara},\ and\ \citenamefont
  {Atat{\"{u}}re}}]{Knowles2014}%
  \BibitemOpen
  \bibfield  {author} {\bibinfo {author} {\bibfnamefont {H.~S.}\ \bibnamefont
  {Knowles}}, \bibinfo {author} {\bibfnamefont {D.~M.}\ \bibnamefont {Kara}}, \
  and\ \bibinfo {author} {\bibfnamefont {M.}~\bibnamefont {Atat{\"{u}}re}},\
  }\href {\doibase 10.1038/nmat3805} {\bibfield  {journal} {\bibinfo  {journal}
  {Nature Materials}\ }\textbf {\bibinfo {volume} {13}},\ \bibinfo {pages} {21}
  (\bibinfo {year} {2014})}\BibitemShut {NoStop}%
\bibitem [{\citenamefont {Ohno}\ \emph {et~al.}(2012)\citenamefont {Ohno},
  \citenamefont {{Joseph Heremans}}, \citenamefont {Bassett}, \citenamefont
  {Myers}, \citenamefont {Toyli}, \citenamefont {{Bleszynski Jayich}},
  \citenamefont {Palmstr{\o}m},\ and\ \citenamefont {Awschalom}}]{Ohno2012}%
  \BibitemOpen
  \bibfield  {author} {\bibinfo {author} {\bibfnamefont {K.}~\bibnamefont
  {Ohno}}, \bibinfo {author} {\bibfnamefont {F.}~\bibnamefont {{Joseph
  Heremans}}}, \bibinfo {author} {\bibfnamefont {L.~C.}\ \bibnamefont
  {Bassett}}, \bibinfo {author} {\bibfnamefont {B.~A.}\ \bibnamefont {Myers}},
  \bibinfo {author} {\bibfnamefont {D.~M.}\ \bibnamefont {Toyli}}, \bibinfo
  {author} {\bibfnamefont {A.~C.}\ \bibnamefont {{Bleszynski Jayich}}},
  \bibinfo {author} {\bibfnamefont {C.~J.}\ \bibnamefont {Palmstr{\o}m}}, \
  and\ \bibinfo {author} {\bibfnamefont {D.~D.}\ \bibnamefont {Awschalom}},\
  }\href {\doibase 10.1063/1.4748280} {\bibfield  {journal} {\bibinfo
  {journal} {Applied Physics Letters}\ }\textbf {\bibinfo {volume} {101}},\
  \bibinfo {pages} {082413} (\bibinfo {year} {2012})}\BibitemShut {NoStop}%
\bibitem [{\citenamefont {Yamamoto}\ \emph {et~al.}(2013)\citenamefont
  {Yamamoto}, \citenamefont {Umeda}, \citenamefont {Watanabe}, \citenamefont
  {Onoda}, \citenamefont {Markham}, \citenamefont {Twitchen}, \citenamefont
  {Naydenov}, \citenamefont {McGuinness}, \citenamefont {Teraji}, \citenamefont
  {Koizumi}, \citenamefont {Dolde}, \citenamefont {Fedder}, \citenamefont
  {Honert}, \citenamefont {Wrachtrup}, \citenamefont {Ohshima}, \citenamefont
  {Jelezko},\ and\ \citenamefont {Isoya}}]{Yamamoto2013}%
  \BibitemOpen
  \bibfield  {author} {\bibinfo {author} {\bibfnamefont {T.}~\bibnamefont
  {Yamamoto}}, \bibinfo {author} {\bibfnamefont {T.}~\bibnamefont {Umeda}},
  \bibinfo {author} {\bibfnamefont {K.}~\bibnamefont {Watanabe}}, \bibinfo
  {author} {\bibfnamefont {S.}~\bibnamefont {Onoda}}, \bibinfo {author}
  {\bibfnamefont {M.~L.}\ \bibnamefont {Markham}}, \bibinfo {author}
  {\bibfnamefont {D.~J.}\ \bibnamefont {Twitchen}}, \bibinfo {author}
  {\bibfnamefont {B.}~\bibnamefont {Naydenov}}, \bibinfo {author}
  {\bibfnamefont {L.~P.}\ \bibnamefont {McGuinness}}, \bibinfo {author}
  {\bibfnamefont {T.}~\bibnamefont {Teraji}}, \bibinfo {author} {\bibfnamefont
  {S.}~\bibnamefont {Koizumi}}, \bibinfo {author} {\bibfnamefont
  {F.}~\bibnamefont {Dolde}}, \bibinfo {author} {\bibfnamefont
  {H.}~\bibnamefont {Fedder}}, \bibinfo {author} {\bibfnamefont
  {J.}~\bibnamefont {Honert}}, \bibinfo {author} {\bibfnamefont
  {J.}~\bibnamefont {Wrachtrup}}, \bibinfo {author} {\bibfnamefont
  {T.}~\bibnamefont {Ohshima}}, \bibinfo {author} {\bibfnamefont
  {F.}~\bibnamefont {Jelezko}}, \ and\ \bibinfo {author} {\bibfnamefont
  {J.}~\bibnamefont {Isoya}},\ }\href {\doibase 10.1103/PhysRevB.88.075206}
  {\bibfield  {journal} {\bibinfo  {journal} {Physical Review B}\ }\textbf
  {\bibinfo {volume} {88}},\ \bibinfo {pages} {075206} (\bibinfo {year}
  {2013})}\BibitemShut {NoStop}%
\bibitem [{\citenamefont {Tetienne}\ \emph {et~al.}(2018)\citenamefont
  {Tetienne}, \citenamefont {de~Gille}, \citenamefont {Broadway}, \citenamefont
  {Teraji}, \citenamefont {Lillie}, \citenamefont {McCoey}, \citenamefont
  {Dontschuk}, \citenamefont {Hall}, \citenamefont {Stacey}, \citenamefont
  {Simpson},\ and\ \citenamefont {Hollenberg}}]{Tetienne2018a}%
  \BibitemOpen
  \bibfield  {author} {\bibinfo {author} {\bibfnamefont {J.-P.}\ \bibnamefont
  {Tetienne}}, \bibinfo {author} {\bibfnamefont {R.~W.}\ \bibnamefont
  {de~Gille}}, \bibinfo {author} {\bibfnamefont {D.~A.}\ \bibnamefont
  {Broadway}}, \bibinfo {author} {\bibfnamefont {T.}~\bibnamefont {Teraji}},
  \bibinfo {author} {\bibfnamefont {S.~E.}\ \bibnamefont {Lillie}}, \bibinfo
  {author} {\bibfnamefont {J.~M.}\ \bibnamefont {McCoey}}, \bibinfo {author}
  {\bibfnamefont {N.}~\bibnamefont {Dontschuk}}, \bibinfo {author}
  {\bibfnamefont {L.~T.}\ \bibnamefont {Hall}}, \bibinfo {author}
  {\bibfnamefont {A.}~\bibnamefont {Stacey}}, \bibinfo {author} {\bibfnamefont
  {D.~A.}\ \bibnamefont {Simpson}}, \ and\ \bibinfo {author} {\bibfnamefont
  {L.~C.~L.}\ \bibnamefont {Hollenberg}},\ }\href {\doibase
  10.1103/PhysRevB.97.085402} {\bibfield  {journal} {\bibinfo  {journal}
  {Physical Review B}\ }\textbf {\bibinfo {volume} {97}},\ \bibinfo {pages}
  {085402} (\bibinfo {year} {2018})}\BibitemShut {NoStop}%
\bibitem [{\citenamefont {Sedlacek}\ \emph {et~al.}(2018)\citenamefont
  {Sedlacek}, \citenamefont {Stuart}, \citenamefont {Slichter}, \citenamefont
  {Bruzewicz}, \citenamefont {Mcconnell}, \citenamefont {Sage},\ and\
  \citenamefont {Chiaverini}}]{Sedlacek2018}%
  \BibitemOpen
  \bibfield  {author} {\bibinfo {author} {\bibfnamefont {J.~A.}\ \bibnamefont
  {Sedlacek}}, \bibinfo {author} {\bibfnamefont {J.}~\bibnamefont {Stuart}},
  \bibinfo {author} {\bibfnamefont {D.~H.}\ \bibnamefont {Slichter}}, \bibinfo
  {author} {\bibfnamefont {C.~D.}\ \bibnamefont {Bruzewicz}}, \bibinfo {author}
  {\bibfnamefont {R.}~\bibnamefont {Mcconnell}}, \bibinfo {author}
  {\bibfnamefont {J.~M.}\ \bibnamefont {Sage}}, \ and\ \bibinfo {author}
  {\bibfnamefont {J.}~\bibnamefont {Chiaverini}},\ }\href {\doibase
  10.1103/PhysRevA.98.063430} {\bibfield  {journal} {\bibinfo  {journal}
  {Physical Review A}\ }\textbf {\bibinfo {volume} {98}},\ \bibinfo {pages}
  {063430} (\bibinfo {year} {2018})}\BibitemShut {NoStop}%
\bibitem [{\citenamefont {Toyli}\ \emph {et~al.}(2010)\citenamefont {Toyli},
  \citenamefont {Weis}, \citenamefont {Fuchs}, \citenamefont {Schenkel},\ and\
  \citenamefont {Awschalom}}]{Toyli2010}%
  \BibitemOpen
  \bibfield  {author} {\bibinfo {author} {\bibfnamefont {D.~M.}\ \bibnamefont
  {Toyli}}, \bibinfo {author} {\bibfnamefont {C.~D.}\ \bibnamefont {Weis}},
  \bibinfo {author} {\bibfnamefont {G.~D.}\ \bibnamefont {Fuchs}}, \bibinfo
  {author} {\bibfnamefont {T.}~\bibnamefont {Schenkel}}, \ and\ \bibinfo
  {author} {\bibfnamefont {D.~D.}\ \bibnamefont {Awschalom}},\ }\href {\doibase
  10.1021/nl102066q} {\bibfield  {journal} {\bibinfo  {journal} {Nano Letters}\
  }\textbf {\bibinfo {volume} {10}},\ \bibinfo {pages} {3168} (\bibinfo {year}
  {2010})}\BibitemShut {NoStop}%
\bibitem [{\citenamefont {de~Wit}\ \emph {et~al.}(2018)\citenamefont {de~Wit},
  \citenamefont {Welker}, \citenamefont {de~Voogd},\ and\ \citenamefont
  {Oosterkamp}}]{DeWit2018}%
  \BibitemOpen
  \bibfield  {author} {\bibinfo {author} {\bibfnamefont {M.}~\bibnamefont
  {de~Wit}}, \bibinfo {author} {\bibfnamefont {G.}~\bibnamefont {Welker}},
  \bibinfo {author} {\bibfnamefont {J.}~\bibnamefont {de~Voogd}}, \ and\
  \bibinfo {author} {\bibfnamefont {T.}~\bibnamefont {Oosterkamp}},\ }\href
  {\doibase 10.1103/PhysRevApplied.10.064045} {\bibfield  {journal} {\bibinfo
  {journal} {Physical Review Applied}\ }\textbf {\bibinfo {volume} {10}},\
  \bibinfo {pages} {064045} (\bibinfo {year} {2018})}\BibitemShut {NoStop}%
\bibitem [{\citenamefont {Friedrich}(2017)}]{Friedrich2017}%
  \BibitemOpen
  \bibfield  {author} {\bibinfo {author} {\bibfnamefont {H.}~\bibnamefont
  {Friedrich}},\ }\href {\doibase 10.1007/978-3-319-47769-5} {\emph {\bibinfo
  {title} {{Theoretical Atomic Physics}}}},\ Graduate Texts in Physics\
  (\bibinfo  {publisher} {Springer International Publishing},\ \bibinfo
  {address} {Cham},\ \bibinfo {year} {2017})\ pp.\ \bibinfo {pages}
  {232--235}\BibitemShut {NoStop}%
\end{thebibliography}%

\clearpage
\onecolumngrid
\subsection*{\Large Supplemental Material}
\normalsize

\setcounter{equation}{0}
\setcounter{figure}{0}
\setcounter{table}{0}
\makeatletter
\renewcommand{\theequation}{S\arabic{equation}}
\renewcommand{\thefigure}{S\arabic{figure}}
\renewcommand{\thetable}{S\arabic{table}}

\section{supplemental note 1: Sample preparation and details}

The diamond substrate used in this work is a 150-$\mu$m-thick plate with optically resolvable near-surface NV centers (depth $\sim 7$ nm). This same substrate is also used and described in Refs.~\cite{Myers2017b,Ariyaratne2018b,Bluvstein2019}. Optical access in the confocal setup is through the 150-$\mu$m thick diamond plate, and a metal waveguide patterned on the diamond is used to transmit microwaves.

The diamond begins as a polished, commercial Element Six electronic grade (100) diamond substrate of lateral dimensions 2 mm $\times$ 2 mm and a thickness of 0.5 mm. To reduce the thickness, the diamond is sliced and then polished from the sliced side to a thickness of 150 $\mu$m. Polishing damage is then mitigated by etching 1 $\mu$m with ArCl$_2$ plasma, and the substrate is then cleaned for 40 minutes in boiling acid H$_2$NO$_3$:H$_2$SO$_4$ 2:3. A 50-nm-thick layer of 99.99\% $^{12}$C is then grown on the diamond face by plasma-enhanced chemical vapor deposition. To form NV centers, $^{14}$N ions are then implanted with a dosage of $5.2\times 10^{10}$ ions/cm$^2$ at 4 keV and a 7$\degree$ tilt, yielding an expected N depth of 7 nm (calculated by Stopping and Range of Ions in Matter (SRIM)). Subsequently, the sample is annealed in vacuum ($< 10^{-6}$ Torr at max temperature) at 850$\degree$ C for 2.5 hours with a 40-minute temperature ramp. After annealing, the sample is cleaned in HClO$_4$:H$_2$NO$_3$:H$_2$SO$_4$ 1:1:1 for 1 hour at 230-240 $\degree$C. Materials processing techniques similar to those used here are detailed and characterized in Reference~\cite{Sangtawesin2018}.

Approximately three years span between the initial sample preparation and the measurements performed in this work, during which time several rounds of surface cleaning are performed. The measurements performed in this work are performed several months after a standard surface preparation protocol: the diamond is acid cleaned with Nanostrip (a Piranha analog) at 80$^{\circ}$C for 12 minutes, and then the diamond is oxygen annealed at 400$^{\circ}$C for 4 hours. Nanostrip is chosen because it preserves the metal waveguide on the diamond.

The NV centers' depths are measured by proton NMR and range between $\sim$4 and 17 nm; further details are given in the supplement of Ref.~\cite{Bluvstein2019}. To increase the photon collection efficiency from the NV centers, tapered nanopillars with a diameter of 400-nm are patterned by e-beam lithography and etched by O$_2$ plasma to a height of 500 nm. Under conventional spin-dependent photoluminescence readout, we measure spin contrasts of $\approx$ 35\% and a saturation fluorescence of $\approx$ 500 k counts/s. Pillars containing single NV centers are identified by second-order correlation measurements and then confirmed by measuring photon statistics from the negative and neutral NV charge states (see supplement of Ref.~\cite{Bluvstein2019}). By measuring the fraction of pillars hosting a single NV center and assuming the number of NVs per pillar follows a Poisson distribution, we estimate a conversion efficiency of implanted N to NV of $\approx 5\%$, consistent with previous reports \cite{Toyli2010}.

On day-to-month timescales, on some NVs we observe order-unity variations in the NV coherence time as well as the NV coupling strength to the surface spins. Accordingly, for Figs.~2 and 3(c) of the main text we interleave drive and no-drive sequences so that we are insensitive to sub-kHz changes in the environment. We note that for the DEER data in Figs.~1(c) and 1(d) of the main text, the NV $T_2$ was slightly lower during the measurement in Fig.~1(c), and so the quantitative values of NV coherence in Fig.~1 should not be directly compared.

\section{Supplemental Note 2: Surface spin density estimate}
Here we describe an estimate of the surface spin density by using the measurements presented in this work. One may use the DEER coupling strength to estimate the surface spin density; however, depending on the models and assumptions applied (\textit{e.g.} about correlation times of the noise), similar DEER coupling strengths can result in estimated densities that vary by orders of magnitude. Accordingly, in this section we present statistical arguments to set a lower bound on the surface spin density of $\approx 0.01 / \text{nm}^2$, and we use knowledge of the surface spin linewidth to set an upper bound of $\approx 0.1 / \text{nm}^2$. The bounded range of $0.01 - 0.1 /\text{nm}^2$ is consistent with the densities extracted by various different methods in Refs.~\cite{Sushkov2014,Myers2014,DeWit2018}.

\subsection*{2.1: Lower bound on surface spin density: statistical argument}

We can set a lower bound on the surface spin density by using statistical arguments, based on the sample of NVs we measure. A low spin density will lead to large variations in the surface spin signal that the individual NV centers feel: if we assume the extreme scenario that all observed variation between different NVs is due to a variation in the number of surface spins in the NV sensing volume, then we can estimate a density of surface spins which would be a lower bound. 

Figure~3(c) of the main text plots $\Gamma_\text{decoupled} = 1/T_{2} - 1/T_{2,\text{drive}}$ for eight individual centers of different depths. Six of the centers are between depths of approximately 7.5 to 12.5 nm; for these six centers, the mean decoupling rate mean($\Gamma_\text{decoupled}$) = 7.75 ms$^{-1}$ and the standard deviation std($\Gamma_\text{decoupled}$) = 2.9 ms$^{-1}$. From the plot, it appears a majority of the variation for these six centers originates from their varied depth, but for this estimate we assume all variation in $\Gamma_\text{decoupled}$ is due to the variation in the number of surface spins in the NV sensing volume.

For this estimate we assume that $\Gamma_\text{decoupled}$ is proportional to the mean-squared magnetic field $B^2_\text{rms}$ from the surface spins. $B^2_\text{rms}$ is proportional to $N$, where $N$ is the number of surface spins in the sensing volume, which here we assume follows a Poisson distribution. Then std($\Gamma_\text{decoupled}$)/mean($\Gamma_\text{decoupled}$) = std($B^2_\text{rms}$)/mean($B^2_\text{rms}$) = $\sqrt{N} /N = 1 /\sqrt{N}$. We observe std($\Gamma_\text{decoupled}$)/mean($\Gamma_\text{decoupled}$) $\approx 0.37$, and accordingly $N \approx 1/(0.37)^2 \approx 7$. Approximately, the relevant sensing area on the surface is a circle with area $\sim \pi r^2$, where $r \approx 10$ nm is the average depth of the six NV centers we consider here. These estimates correspond to a surface spin density of $7 / \pi (10 ~\text{nm})^2 \approx 0.02 / \text{nm}^2$, and to be conservative we estimate a lower bound of $\approx 0.01 / \text{nm}^2$.

This argument becomes invalid if the surface spins are not stationary over the course of the entire measurement of $\Gamma_\text{decoupled}$ on an NV, which lasts minutes to hours. If the spins are not stationary, then an arbitrarily low spin density would not necessarily result in any variation between different NV centers. However, the spins are likely stationary on the surface, as Sushkov \textit{et al}. \cite{Sushkov2014} image the locations of these surface spins by varying the magnetic field direction.

\subsection*{2.2: Upper bound on surface spin density: dipolar-limited surface spin linewidth}

We can set an upper bound on the surface spin density by noting the typical surface spin linewidths we observe, because dipolar coupling between the surface spins will broaden the line. In this work, we observe a surface spin $1/T_{2,\text{Rabi}} \approx 5$ MHz, roughly corresponding to the half width at half maximum (HWHM) of the resonance linewidth. Note that HWHMs of approximately 10 MHz are observed in Figs.~1(c) and 3(b) of the main text, but both of these HWHMs are Fourier broadened by approximately 5 MHz. If we assume that the measured 5 MHz decay constant is exclusively due to dipolar interactions between the surface spins, then we can set an upper bound on the surface spin density.

Roughly, the linewidth will be of the order of the dipolar coupling strength $A$. Consider, for example, the case of two spins with dipolar interaction strength $A$ and Zeeman interaction energy $\gamma B$, where $\gamma B \gg A$. The transition frequency of $|\downarrow \downarrow \rangle \rightarrow|\uparrow \downarrow \rangle$ will be $\gamma B - A$, and the transition frequency of $|\downarrow \uparrow \rangle \rightarrow|\uparrow \uparrow \rangle$ will be $\gamma B + A$. Having large numbers of dipolar-coupled spins thus creates a broadening with a HWHM of order $A$. Slichter calculates the second moment (\textit{i.e.} the variance) of the linewidth given $N$ spins limited by dipolar broadening \cite{Slichter1990}, and this variance is given by

\begin{equation}
\langle\Delta f^2\rangle = \left(\frac{1}{2\pi}\right)^2 \left(\frac{\mu_0}{4\pi}\right)^2\frac{3}{4} \gamma^4 \hbar^2 S(S+1) \frac{1}{N} \sum_{j,k} \frac{(1-3\cos^2{\theta_{jk}})^2}{r^6_{jk}},
\label{slichter}
\end{equation}

where $\gamma$ is the electron gyromagnetic ratio, $S$ is the spin (1/2 for the spin-1/2 surface spins), $N$ is the number of spins, and the sum adds the contribution from all spins, which are at a relative separation $r_{jk}$ and point at a relative angle $\theta_{jk}$.

As in Slichter \cite{Slichter1990}, we now consider the case where all spins are located in equivalent positions, such that the sum is independent of $j$. This then gives $N$ equivalent sums, canceling with the $1/N$. We arrive at

\begin{equation}
\Delta f_\text{rms} \equiv \sqrt{\langle\Delta f^2\rangle} \approx \text{39 MHz nm$^3$} \sqrt{\sum_{k} \frac{(1-3\cos^2{\theta_{k}})^2}{r^6_{k}}}.
\label{slichter2}
\end{equation}

We approximate the sum for our surface spins by considering a two-dimensional square array of spins. We use the nearest neighbor approximation (\textit{i.e.}, we sum the contribution from the four nearest neighbors to the central spin), and we consider a spin z-axis which is tilted 54.7$^{\circ}$ from the axis normal to the array (the direction of the applied magnetic field for our (100) diamond). These considerations then yield an upper bound for the surface spin density given by

\begin{equation}
\sigma \approx \frac{1}{\text{nm}^2} \left(\frac{\Delta f_\text{rms}}{125~ \text{MHz}}\right)^{2/3}.
\label{sigmahigh}
\end{equation}

If we take $\Delta f_\text{rms}$ to be our experimentally measured HWHM of 5 MHz, then Eq.~\ref{sigmahigh} gives an upper bound to the surface spin density of $\approx 0.1 / \text{nm}^2$.We emphasize that, by our calculations here, a surface spin density of $1/\text{nm}^2$ would give a FWHM of 250 MHz, far inconsistent with the observed linewidths in this work and in previous reports \cite{Mamin2012,Grinolds2014,Sushkov2014}.

This argument assumes that the spins lie perfectly in a 2D plane, without any height variation: however, rapid nm changes in height over nm lengthscales would lead to lower average dipolar coupling strengths than if the spins are all in the same plane. We also note that one should be careful in the interpretation of linewidth. A Lorentzian lineshape, for example, has infinite standard devation but a well-defined HWHM. In the case of the two-dimensional square array of spins, the HWHM is approximately the same value as the standard deviation and so we can use the method of moments safely.

\section{supplemental note 3: Radio-frequency control}

\subsection*{3.1: Radio-frequency equipment}
Three separate radio-frequency (RF) signal generators are used for controlling the electronic spin states of the NV spin qutrit, formed by $|m_s=0\rangle_{\text{NV}}$ and $|m_s=\pm1\rangle_{\text{NV}}$, as well as the surface spin qubits, formed by $|\uparrow \rangle_{\text{SS}}$ and $|\downarrow\rangle_{\text{SS}}$. A single microwave waveguide patterned on the diamond is used to deliver all RF signals. All three RF generators (SRS SG384) are gated by their own switch (Mini Circuits ZASWA-2-50DR+) and are then combined before an RF amplifier (Mini Circuits ZHL-16W-43+) which feeds into the patterned waveguide. Independent IQ modulation of each SG384 generator is used as an extra layer of isolation. Measurements are done in an applied magnetic field of 100s of Gauss, and so the surface spin RF frequency is $\sim$ GHz and passes through the amplifier.

\subsection*{3.2: Preparation, control, and readout of double-quantum basis}

The $|m_s=-1\rangle_{\text{NV}}$ and $|m_s=+1\rangle_{\text{NV}}$ states are not coupled by microwaves, so to manipulate superpositions in the DQ basis we employ microwave control of the SQ bases $\{0,-1\}_{\text{NV}}$ and $\{0,+1\}_{\text{NV}}$. We use subsequent single-frequency pulses to prepare, control, and readout DQ superpositions. For example, to prepare a DQ superposition, we apply an $|m_s=0\rightarrow-1\rangle_{\text{NV}} ~\pi/2$ pulse to prepare a SQ superposition, and then apply an $|m_s=0\rightarrow+1\rangle_{\text{NV}} ~\pi$ pulse. The DQ Hahn echo sequence we use in Fig.~2 of the main text is depicted in Fig.~\ref{DQecho}, which includes use of a differential measurement (see below). Dual-frequency pulses have been shown to also be an effective way to control the DQ basis \cite{Mamin2014}, at the cost of experimental complexity; however, dual-frequency pulses would likely be superior for use in measurements with long trains of RF pulses, such as CPMG-type measurements.

\subsection*{3.3: Differential measurement}
For all measurements plotted in the main text and supplement of this work, we use a common-mode rejection technique known as a differential measurement, used throughout the NV literature and described in Refs.~\cite{Myers2017b,Ariyaratne2018b,Bluvstein2019}. The differential measurement helps remove technical noise, can alleviate some effects of NV charge state conversion in the dark \cite{Bluvstein2019}, and also simplifies analysis by normalizing the signal so that the signal goes to $0$ as $\tau \rightarrow \infty$ in an echo sequence.

\begin{figure}
\includegraphics[width=152mm]{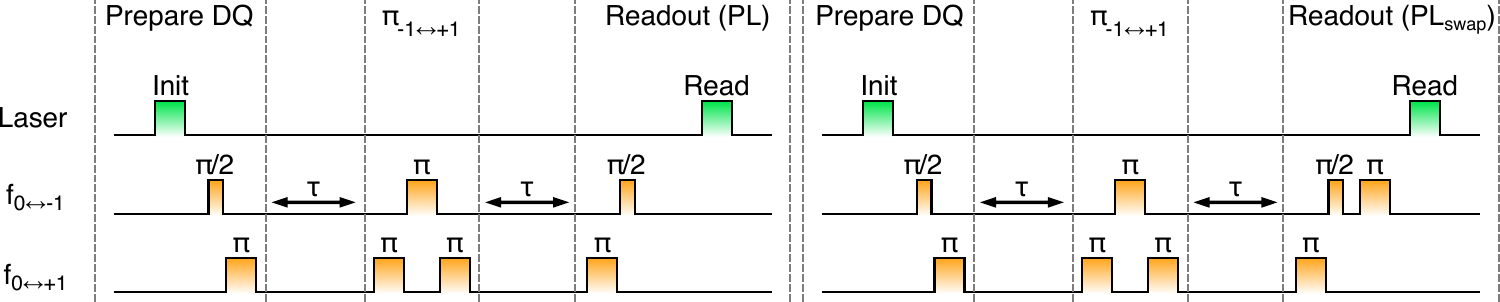}
\caption{Pulse sequence used for performing Hahn echo measurements in the double-quantum basis, which includes use of a differential measurement technique by the additional $\pi$ pulse at the end of the $\text{PL}_\text{swap}$ measurement. The NV coherence is proportional to the differential measurement signal $\text{PL} - \text{PL}_\text{swap}$.
}
\label{DQecho}
\end{figure}

In the differential measurement scheme, at the end of a pulse sequence we readout the NV photoluminescence, denoted PL in Fig.~\ref{DQecho}. We then repeat the pulse sequence, but immediately before photoluminescence readout we apply a resonant microwave $\pi$ pulse to swap the $|m_s=0\rangle_{\text{NV}}$ and $|m_s=-1\rangle_{\text{NV}}$ populations and measure PL$_\text{swap}$ (Fig.~\ref{DQecho}). Subtracting PL and PL$_\text{swap}$ thus removes background PL signals and yields the population difference between $|m_s=0\rangle_{\text{NV}}$ and $|m_s=-1\rangle_{\text{NV}}$.
Explicitly, the overall signal we measure is given by 

\begin{equation}
\text{PL} - \text{PL}_\text{swap} = C \left(\rho_0 - \rho_{-1}\right),
\label{Diff}
\end{equation}

where $C$ is a constant describing the spin-dependent PL contrast, and $\rho_0$ and $\rho_{-1}$ are the $|m_s=0\rangle_{\text{NV}}$ and $|m_s=-1\rangle_{\text{NV}}$ populations after the projective $\pi/2$ pulse. Throughout this work, we define NV coherence as $\left(\text{PL} - \text{PL}_\text{swap}\right)/C$.

\section{supplemental note 4: The microwave analog of the AC Stark effect}

In this section we show that the drive field used for addressing the surface spins can cause an appreciable shift in the NV spin transition frequencies, even if far detuned from the NV spin transition. We identify this effect as a microwave analog of the AC stark effect, common to optical transitions in atomic physics. We observe this effect in our DEER measurement sequences as a slow oscillation in the NV phase as a function of the surface spin pulse duration, and we show in this section that this effect can be readily removed with the appropriate fitting function. We note that we also observe this effect as a frequency shift in Ramsey experiments (not plotted in this work). We emphasize that these results indicate that amplitude fluctuations in the surface spin drive field, even though far detuned from the NV transition, would lead to NV dephasing. Similarly, spatial inhomogeneity in the drive field would lead to dephasing in NV ensemble measurements.

\subsection*{4.1: Theoretical description of AC Stark effect on NV spin triplet}

The AC Stark effect is common to optical transitions in Atomic physics, and is described in detail in \textit{Theoretical Atomic Physics} by Friedrich \cite{Friedrich2017}. Here we adapt the calculation to the magnetic transitions of the NV ground state spin triplet. To be explicit, the NV interaction with the Rabi field can be understood by expanding the Zeeman interaction:

\begin{equation}
H^{\prime} = \gamma\boldsymbol{S} \cdot \boldsymbol{B} \cos(\omega t) = \left[\gamma B_z S_z + \left(\frac{1}{2} \gamma \left(B_x - i B_y\right) S_+ + \frac{1}{2} \gamma \left(B_x + i B_y\right) S_-\right)\right]\cos(\omega t),
\label{Stark1}
\end{equation}

where $\gamma$ is the electron gyromagnetic ratio, $\boldsymbol{B}$ is the magnetic field amplitude vector, $\cos(\omega t)$ describes the oscillation of the magnetic field, $\boldsymbol{S}$ is the spin-1 operator, and $S_{\pm} = S_x \pm i S_y$ are the spin raising and lowering operators. Analagous to the calculation in Friedrich \cite{Friedrich2017}, the energy shifts (up to second-order) to the NV spin triplet states due to an oscillating magnetic field are thus given by

\begin{equation}
\begin{split}
\Delta E_n &= \frac{1}{4} \left[ \sum_{E_m + \hbar \omega \neq E_n} \frac{|\langle\psi_n|\gamma\boldsymbol{S} \cdot \boldsymbol{B}|\psi_m\rangle|^2}{E_n - E_m - \hbar \omega} \right. \\ &+ \left. \sum_{E_m - \hbar \omega \neq E_n} \frac{|\langle\psi_n|\gamma\boldsymbol{S} \cdot \boldsymbol{B}|\psi_m\rangle|^2}{E_n - E_m + \hbar \omega} \vphantom{\sum_{E_m + \hbar \omega \neq E_n}} \right],
\label{Stark2}
\end{split}
\end{equation}

where $\psi_n$ are the NV $m_s = 0,+1,-1$ spin states, and $E_n$ is the energy of the $n^{\text{th}}$ level. Expanding the Zeeman interaction we arrive at

\begin{equation}
\begin{split}
\Delta E_n &= \frac{1}{4} \left[ \sum_{E_m + \hbar \omega \neq E_n} \frac{|\langle\psi_n|\left(\frac{1}{2} \gamma \left(B_x - i B_y\right) S_+ + \frac{1}{2} \gamma \left(B_x + i B_y\right) S_-\right)|\psi_m\rangle|^2}{E_n - E_m - \hbar \omega} \right. \\ &+ \left. \sum_{E_m - \hbar \omega \neq E_n} \frac{|\langle\psi_n|\left(\frac{1}{2} \gamma \left(B_x - i B_y\right) S_+ + \frac{1}{2} \gamma \left(B_x + i B_y\right) S_-\right)|\psi_m\rangle|^2}{E_n - E_m + \hbar \omega} \vphantom{\sum_{E_m + \hbar \omega \neq E_n}} \right],
\label{Stark3}
\end{split}
\end{equation}

where the $S_z$ term is cancelled out because the inner product will be $0$ unless $n=m$ for $S_z$, but if $n=m$ then the two terms cancel. In our experiment, we are interested in the relative energy shift induced between two spin levels. Here we will focus in particular on the relative energy shift between $m_s = 0$ and $m_s = -1$, which we use as our single-quantum qubit. Calculation with Eq.~\ref{Stark3} results in

\begin{equation}
\hbar\omega_{\text{stark}} \equiv
\Delta E_{-1} - \Delta E_{0} = \hbar \Omega^2_\text{NV} \left(\frac{1}{2\Delta_{-1}} + \frac{1}{2\Delta_{-1} + 4 \omega} + \frac{1}{4\Delta_{+1}} + \frac{1}{4\Delta_{+1} + 8 \omega}
\right)
\label{Eshift}
\end{equation}

where $\Omega_\text{NV} = (1/\sqrt{2})\gamma \sqrt{B^2_x + B^2_y}$ describes the NV coupling to the Rabi field, and $\Delta_{\pm1} = \omega_{\pm 1} - \omega$ are the detunings of the drive frequency $\omega$ from the $|m_s = 0 \rightarrow \pm 1 \rangle_{\text{NV}}$ transition frequencies $\omega_{\pm 1}$. Note that for $\Delta_{-1} \ll \omega, \Delta_{+1}$, Eq.~\ref{Eshift} becomes approximately $\hbar \Omega^2_\text{NV} / 2\Delta_{-1}$, which is commonly quoted as the AC Stark shift.

In this work we measure $\Omega_\text{SS}$ as a proxy for $\Omega_\text{NV}$, where $\Omega_\text{SS}$ is the surface spin coupling to the Rabi field. We emphasize that $\Omega_\text{NV} = \sqrt{2} ~\Omega_\text{SS}$, assuming the magnetic field is aligned to the NV axis. This $\sqrt{2}$ factor is due to the nature of the NV being a spin-1 particle, compared to the surface spins which are spin-1/2. Explicitly, the extra factor of $\sqrt{2}$ arises because $S_+ |S=\frac{1}{2}, m_s = -\frac{1}{2} \rangle = \hbar |S=\frac{1}{2}, m_s = +\frac{1}{2} \rangle$, whereas $S_+ |S= 1, m_s = -1 \rangle = \sqrt{2}~ \hbar |S=1, m_s = 0 \rangle$.

In summary, the theoretical result derived in this section shows that a relative energy shift is induced between the NV spin levels due to an off-resonant RF drive. Under the experimental parameters associated with the drive field used to address the surface electron spins in this work, this energy shift can be appreciable. For example, even at a large detuning of $\Delta_{-1}/2\pi =$ 1000 MHz, a drive field of strength $\Omega_\text{NV}/2\pi = 20$ MHz yields a Stark shift of approximately $(1/2\pi)~ \Omega^2_\text{NV} / 2\Delta_{-1} = 0.2$ MHz, which is easily observable with an NV $T_2$ of 10-100 microseconds. Further, these results indicate that the stability of the surface spin drive field is important for maintaining NV coherence. For example, for a 0.2 MHz shift, RF intensity variations of $\sim$ 10\% would lead to $\sim$ 20 kHz variations in the transition frequency and limit the NV coherence to $\sim$ 50 $\mu$s.

\subsection*{4.2: Experimental observation of AC Stark effect and its effect on NV measurements}

\begin{figure}
\includegraphics[width=159mm]{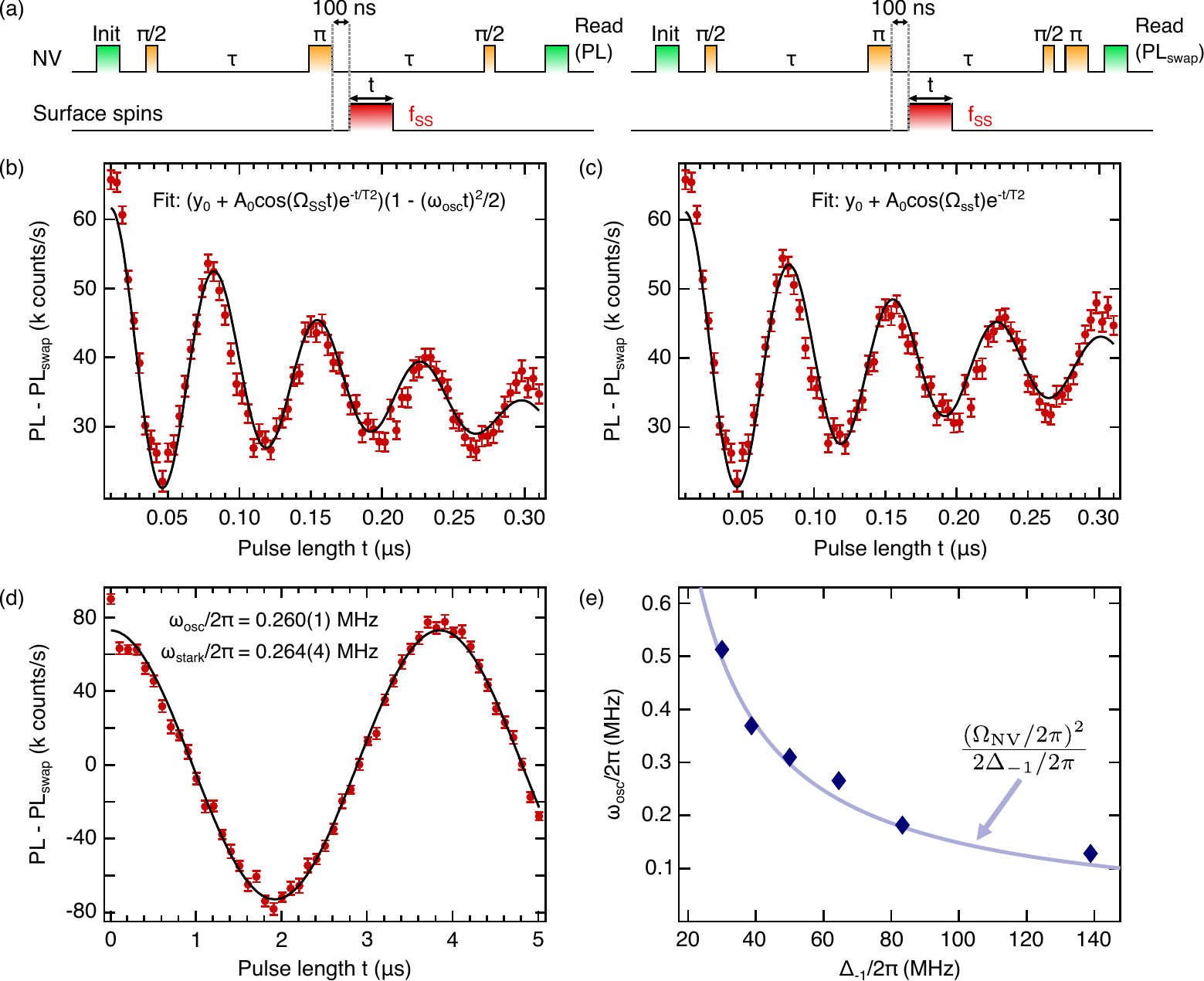}
\caption{Microwave analog of the AC Stark effect and its effect on NV measurements. (a) DEER pulse sequence used in this work. The 100-ns offset is used for technical reasons and does not affect measurements of surface spin properties. (b) Rabi oscillations of surface spins measured by the DEER sequence in (a). The data has a  downward trend caused by the AC Stark effect, which we fit by including an additional factor of $(1 - (\omega_\text{osc} t)^2/2)$, where $\omega_\text{osc}$ is a free fit parameter corresponding to the $\cos(\omega_\text{osc} t)$ signal observed in (d). (c) Data from (b) after dividing by the fitted $(1 - (\omega_\text{osc} t)^2/2)$. In both (b) and (c), the fitted $\Omega_\text{SS} / 2\pi = 13.7(1)$ MHz. (d) Same measurement as (b), with the same experimental parameters but for longer pulse lengths $t$. A coherent phase oscillation that persists far beyond the surface spin coherence time is observed and is fit to $A \cos(\omega_\text{osc} t)$. The fitted oscillation rate $\omega_\text{osc}$ agrees with the predicted oscillation frequency $\omega_\text{stark}$, which is calculated independent of the data in (d). To calculate $\omega_{\text{stark}}$ with Eq.~\ref{Eshift}, we use the experimental values: the applied drive field is $\omega/2\pi$ = 790 MHz, the state detunings are $\Delta_{-1} / 2\pi = 1292$ MHz and $\Delta_{+1} / 2\pi = 2866$ MHz, and the NV coupling to the Rabi field is $\Omega_\text{NV} / 2\pi = \sqrt{2}~ \Omega_\text{SS} /2\pi= \sqrt{2} ~13.7(1)$ MHz. (e) Measured $\omega_\text{osc}/2\pi$ for varied detunings $\Delta_{-1}/2\pi$, taken at $\Omega_\text{NV}/2\pi = 5.5$ MHz. The data are consistent with the plotted curve, which is calculated independent of the plotted data (no fit parameters).
}
\label{starkfig}
\end{figure}

In this section, we focus on the AC Stark effects on DEER in particular, and we show that the AC Stark effect can be removed by an appropriate fitting function. Figure~\ref{starkfig}(a) shows the DEER measurement pulse sequence. We start the surface spin pulse 100 ns after the NV $\pi$ pulse ends due to technical reasons, \textit{e.g.} we have observed that sending both pulses at the same time through the RF circuit can cause artefacts which reduce pulse fidelity.

In Fig.~\ref{starkfig}(b) we plot the DEER signal $\text{PL} - \text{PL}_\text{swap}$ (which is proportional to NV coherence, see Supplemental Material Note 3.3) when applying resonant microwaves to the surface spins at frequency $f_\text{SS}$. Rabi oscillations of the surface spins are observed, but there is also a downward trend of the data. At longer pulse lengths $t$, we observe in Fig.~\ref{starkfig}(d) that the downward trend observed in Fig.~\ref{starkfig}(b) is part of a $\cos(\omega_\text{osc} t)$ oscillation in the NV signal due to a coherent phase accumulation at rate $\omega_\text{osc}$. Thus, the downward trend in Fig.~\ref{starkfig}(b) can be removed by dividing by $\cos(\omega_\text{osc} t)$. In practice, we typically fit the DEER signal with the Taylor expanded $(1 - (\omega_\text{osc} t)^2/2)$; dividing this expression from the data yields the plot in Fig.~\ref{starkfig}(c).

The coherent phase accumulation observed in Fig.~\ref{starkfig}(d) is due to the AC Stark effect. We fit the data in Fig.~\ref{starkfig}(d) to $A\cos(\omega_\text{osc} t)$, where $A$ is in k counts/s and the fitted $\omega_\text{osc}/2\pi = 0.260(1)$ MHz is the oscillation frequency. Using Eq.~\ref{Eshift}, we calculate that the AC stark effect should yield an oscillation frequency $\omega_{\text{stark}}/2\pi = 0.264(4)$ MHz, consistent with the fitted $\omega_\text{osc}/2\pi = 0.260(1)$ MHz. To calculate $\omega_{\text{stark}}$ with Eq.~\ref{Eshift}, we use the experimental values: the applied drive field is at frequency $\omega/2\pi$ = 790 MHz, the state detunings are $\Delta_{-1} / 2\pi = 1292$ MHz and $\Delta_{+1} / 2\pi = 2866$ MHz, and the NV coupling to the Rabi field is $\Omega_\text{NV} / 2\pi = \sqrt{2}~ \Omega_\text{SS} /2\pi= \sqrt{2} ~13.7(1)$ MHz, where $\Omega_\text{SS}$ is fit in Figs.~\ref{starkfig}(b) and \ref{starkfig}(c). Here we use the surface spin Rabi oscillations in order to measure the NV coupling to the Rabi field, possible because the magnetic field is aligned to within a few degrees of the NV axis. Again, we emphasize the $\sqrt{2}$ factor that arises from careful consideration of the difference between the spin-1 NV and the spin-1/2 surface spins. The agreement here between theory and experiment is thus strong experimental evidence that these surface spins are spin-1/2 particles.

In Fig.~\ref{starkfig}(e) we plot measured values of $\omega_\text{osc}/2\pi$ as a function of drive detuning from the $|m_s=0 \rightarrow -1 \rangle_{\text{NV}}$ transition (here the microwaves are not resonant with the surface spins). At these comparatively small detunings, we find good agreement of $\omega_\text{osc}$ with $\Omega^2_\text{NV} / 2\Delta_{-1}$, where here we measure $\Omega_\text{NV}$ by measuring NV Rabi oscillations on resonance. Figure~\ref{starkfig}(e) plots the average of the measured oscillation rates at positive detuning $\Delta_{-1}$ and negative detuning $\Delta_{-1}$, which renders the measurement first-order insensitive to contributions from the $|m_s=0 \rightarrow +1 \rangle_{\text{NV}}$ transition and changes in $\Omega_\text{NV}$ as a function of frequency (due to nonlinearities in the RF circuit).

\section{supplemental note 5: contributions to NV dephasing}

\subsection*{5.1: Quantitative comparison of SQ and DQ coherence times}

We show here that the ratio of the single-quantum (SQ) and double-quantum (DQ) coherence times observed in Fig.~2(c) of the main text indicates there is significant common-mode noise decoupled by use of the DQ basis \cite{Kim2015}. As discussed in the main text, under a large applied magnetic field $\boldsymbol B = B_z \hat{z}$, the SQ transition frequencies are given by

\begin{equation}
    f_{0\rightarrow\pm1} \approx D + d_{\|} \Pi_{\|} / h \pm \left(\frac{\gamma}{2\pi} B_z + \frac{1}{2} \frac{\left(d_{\perp} \Pi_{\perp}/h\right)^2}{(\gamma/2\pi) B_z}\right),
    \label{SQFrequenciesSupp}
\end{equation}

and the DQ transition frequency is given by

\begin{equation}
    f_{-1\rightarrow+1} \approx 2 \left(\frac{\gamma}{2\pi} B_z + \frac{1}{2} \frac{\left(d_{\perp} \Pi_{\perp}/h\right)^2}{(\gamma/2\pi) B_z}\right).
    \label{DQFrequenciesSupp}
\end{equation}

where $h$ is Planck's constant, $D = 2.87$ GHz is the crystal-field splitting, $B_z$ is the magnetic field magnitude along the NV axis, $\gamma/2\pi = 2.8$ MHz/G is the NV gyromagnetic ratio, $d_{\|}/h = 0.35$ Hz$\cdot$cm/V and $d_{\perp}/h = 17$ Hz$\cdot$cm/V are the components of the NV's electric dipole parallel and perpendicular to the $z$-axis, and $\Pi_{\|}$ and $\Pi_{\perp}$ are the parallel and perpendicular components of the effective electric field, where $\boldsymbol{\Pi} = (\boldsymbol{E} + \boldsymbol{\sigma})$ has both electric field $\boldsymbol{E}$ and appropriately scaled strain $\boldsymbol{\sigma}$ terms. For simplicity, we define $\Tilde{E}$ and $\Tilde{B}$ such that

\begin{equation}
\begin{aligned}
    f_{0\rightarrow\pm1} = \Tilde{E} \pm \Tilde{B}\\ f_{-1\rightarrow+1} = 2 \Tilde{B}.
    \end{aligned}
\label{Freqs}
\end{equation}

Fluctuations in $\Tilde{E}$ and $\Tilde{B}$ lead to fluctuations in accumulated phase between measurements, and hence cause dephasing. As calculated in Ref.~\cite{Kim2015}, for the spin echo this results in coherences $C(T)$ as a function of the phase accumulation time $T$ given by

\begin{equation}
    \begin{aligned}    
    C_{\text{SQ}}(T) &= \exp\left[-\frac{\langle(\delta \phi_{\Tilde{E}})^2\rangle + \langle(\delta \phi_{\Tilde{B}})^2\rangle]}{2}\right] \approx \exp\left[-\left(\frac{T}{T_{2,\text{SQ}}}\right)^{n_\text{SQ}}\right] \\ 
    &C_{\text{DQ}}(T) = \exp\left[-2 \langle( \delta \phi_{\Tilde{B}})^2\rangle \right] \approx \exp\left[-\left(\frac{T}{T_{2,\text{DQ}}}\right)^{n_\text{DQ}}\right],
    \end{aligned}
\label{Coherences}
\end{equation}

where $\langle(\delta \phi_{\Tilde{E}})^2\rangle$ and $\langle(\delta \phi_{\Tilde{B}})^2\rangle$ are the variances in accumulated phase induced by $\Tilde{E}$ and $\Tilde{B}$, $T_{2,\text{SQ}}$ and $T_{2,\text{DQ}}$ are the coherence times of the single- and double-quantum bases, and $n_\text{SQ}$ and $n_\text{DQ}$ are stretch factors. Note that this calculation assumes that $\delta \phi_{\Tilde{E}}$ and $\delta \phi_{\Tilde{B}}$ are uncorrelated Gaussian random variables \cite{Kim2015}. If we assume that the NV is only dephased by magnetic field fluctuations, then $\langle(\delta \phi_{\Tilde{E}})^2\rangle = 0$, and if we assume that $n_\text{SQ} \approx n_\text{DQ} \equiv n$, then we can write

\begin{equation}
\left(\frac{T_{2,\text{DQ}}}{T_{2,\text{SQ}}}\right)^n \approx 0.25, \hspace{10pt}\text{for}~ \langle(\delta\phi_{\Tilde{E}})^2\rangle = 0.
\label{ratio}
\end{equation}

And if $\langle(\delta\phi_{\Tilde{E}})^2\rangle \neq 0$, then $\left(T_{2,\text{DQ}}/T_{2,\text{SQ}}\right)^n > 0.25$. For the data in Fig.~2(c) of the main text, we observe $(T_{2,\text{DQ}} / T_{2,\text{SQ}})^n = 0.48(4)$, where the fitted $n \approx 1.6$. This measured value indicates there is substantial common-mode noise eliminated by use of the double-quantum basis.

\subsection*{5.2: CPMG $T_2$ and Ramsey $T^*_2$}

In addition to Hahn echo measurements, we also perform CPMG-8 and Ramsey sequences with and without surface spin driving. By driving the surface spins we observe a CPMG-8 $T_{2,\text{SQ}}$ increase for three out of three NV centers measured, where the fractional increase on each center is similar to the Hahn echo $T_{2,\text{SQ}}$ increase. Future work can carefully map out the decoupled noise spectrum to evaluate the efficacy of surface spin driving as a function of the number of $\pi$ pulses, in both the SQ and DQ bases.

We do not observe a change in the Ramsey $T^*_{2,\text{SQ}}$ with surface spin driving, which is consistent with the fact that the $1/T^*_{2,\text{SQ}}$ values we measure (100s of kHz) are much greater than the DEER coupling rates we measure (10s of kHz). We attribute the other contributions to $1/T^*_{2,\text{SQ}}$ to low-frequency noise that is independent of the surface spins, such as electric field noise, magnetic noise from P1 centers, and drift of the applied magnetic field ($B_0 \sim 300$ G).

\end{document}